# Hidden variables in Mermin's GHZ machine with quantum-assistance: an introduction to quantum superdeterminism

Jose C. Moreno*

*New American Quantum education Society Operations (NAQESO)*

*Morenj13@alumni.uci.edu*

*Three experiments, with an IBM superconducting quantum computer, are presented, where the setting combinations on a three-qubit GHZ-like state were selected by two additional assistant qubits. The average of Mermin's polynomial for the three entangled qubits was calculated; the results showed "violation" of Mermin's inequality. However, given that the assistant qubits selected, imposed and informed the type of settings, it was possible to interpret the results in terms of arranged relations among "hidden" variables of the assistants and the entanglement BEFORE each "shot"; the "hidden" variables may or may not be local depending on the way the qubits were initialized.*

**Contents**

Mermin's GHZ Machine (MGHZM)

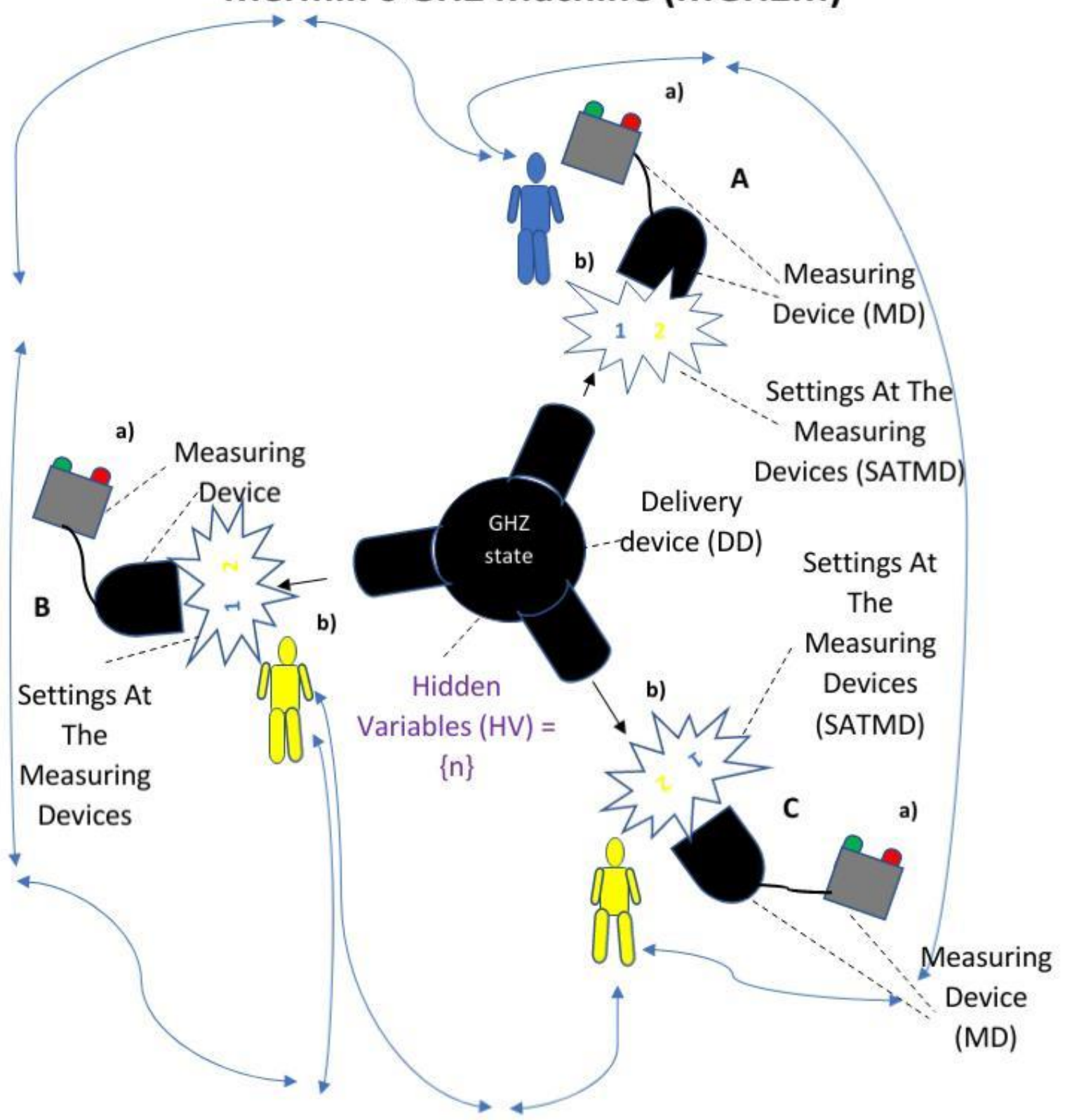

**Figure 1**. **a)** measuring devices at A,B,C flash a non-odd number of reds (i.e odd number of greens) for the setting combinations (1,2,2), (2,1,2), (2,2,1), or an odd number of reds (i.e non-odd number of greens) for the setting combination (1,1,1).**b)** The settings are adjusted strategically at each detector A,B,C, by the scientists, or by some other process, so that each of the setting combinations (1,2,2), (2,1,2), (2,2,1) can occur; suddenly, two of the settings in position 2 (by yellow scientists) are changed while the other remains at setting 1 (by blue scientist) so that they become (1,1,1).

# I. INTRODUCTION

## A. A new type of Mermin's GHZ machine

In "Quantum Mysteries Revisited" [1], David Mermin presents a device, which will be referred in this paper as **Mermin's GHZ machine** (MGHZM), that delivers three quanta to three separate detectors capable of measuring their respective quanta in two possible ways: either all measuring devices were in setting 1, or any two measuring devices had settings 2 while the other had setting 1 (fig. 1). The measuring devices flash either their green or red lights when the quanta arrive such that when all settings are 1, the devices flash an odd number of red flashes (non-odd number of green flashes); otherwise, with the other possible settings, the devices flash a non-odd number of red flashes (odd number of green flashes).[1]

Mermin considers the possibility that there are "**hidden" variables** (HVs), in those quanta that depart from the **delivery device** (DD), which carry "instructions" to flash the **measuring devices** (MD) based on the setting they encounter once they arrived. Those HV are organized in the DD somehow; that is the GHZ state. Mermin begins his analysis by having two of the **settings at the measuring devices** (SATMD) be 2 while the other one was 1; the quanta are assumed to carry "rules" that always result in a non-odd number of red flashes with those **setting combinations** (SCs) (fig 1**a**). Mermin shows that if suddenly, before the quantum information arrived, those in position 2 were switched to the position 1 (fig. 1b), they would not render now an odd number of red flashes as would be expected, disproving the assumption that there exists a general "rule" for the quanta that would give all the correct results. Given that the "rules" can only work for a subset of the possible SCs, and no "rules" in the three quanta could anticipate the **delayed-choice of settings** (DCOS) for the SATMD, it seems futile to believe that there are HVs.

However, there is another possibility if one considers a new method for a DCOS, for the SATMD, assisted by another quantum mechanical system: *"special" SATMD that are able to anticipate the HVs of the three quanta, using the information from **quantum-assistant devices** (QADs), whose HVs are already arranged to impose SCs consistent with the information carried by those three quanta*. Suppose that the SCs, for the three "special" SATMDs, were dependent upon the outcome of the QADs, whose output could flash either blue or yellow, as shown in fig. 2. Let's say that first two of them remain by default set to 2, while a blue flash imposes a 1 on the other; after a brief moment, a blue or yellow flash sets a 1 or keeps the 2, respectively (fig. 2a), on those set to 2 by default. Like in the MGHZM above, when all three settings are 1, the outcome is an odd number of red flashes; when two are set to 2, and only one to 1, the outcome is a non-odd number of red flashes. The QAD directly activates the settings by sending quanta with information necessary to "activate" those SCs. If the outcomes, either blue or yellow, in the QADs, defined the settings before the arrival of the quanta from the DD, the scientists would just know, without having imposed the settings themselves, if all three were going to flash a non-odd or an odd number of reds based on those QADs flashes, and they can evaluate after all measurements were realized whether or not the number of flashes were as expected. In this way, in this version of MGHZM, rather than allowing the scientists to make the "choice" of settings by their gut feeling, or by the assistance of some classical mechanical apparatus that controls the SATMD in all three stations so that they are either 2 or 1, the selections are made by an additional quantum mechanical system where the scientists passively affirm the disclosure from the additional system and can assess if the quantum assisted MGHZM (QAMGHZM) works as expected, implying the SC from the type of flashes from the QAD: *those "special" SATMD can only be implied by the QAD's flashes.* Applying MGHZM analysis to this quantum-assisted thought experiment, one can easily see that the conclusion, that there are no HVs, that can give ALL the results for the SCs (1,2,2),(2,1,2),(2,2,1),(1,1,1)) does not follow if the QAD can coordinate its "rules" with the quantum information from the DD; by "rigging" the HVs from the three quanta in the DD with those HVs in the QAD one can sustain the existence of HVs in both systems without a contradiction, even by DCOS. In other words, when one of the settings is 1 (one blue flash), the other two by default are 2 (no flashes yet from the QAD), and the settings of those set to 2 are suddenly changed to 1 (by QAD flashes), in "mid-flight" (when the quantum information from DD is arriving to the SATMDs), one can interpret that the QAD had the "instructions" already, by flashing blue, so that the settings were consistent with the odd number of red "instructions" that were arriving to the SATMD, precisely to flash an odd number of reds after measurements. Otherwise, if the DD had the "instructions" for non-odd number of reds, the QAD flashes yellow. Now, to test the device systematically, one has to make sure that all three detectors have the opportunity to be set to 2; that is, the QADs could be rotated, as shown in fig. 2b**.**, which would result in four types of settings: (1,2,2), (2,1,2), (2,2,1), and (1,1,1). By doing so, one could test whether or not the HVs of the entanglement are able adjust to the SCs that result in odd number or a non-odd number of red flashes as if the four type of SCs were trying to "fool" the information that arrives at the SATMD. If the machine functions as expected, three of those SCs should result in a non-odd number of red flashes (odd greens), in every repetition of the experiment, and in only one SC, an odd number of reds (even greens). Because in the QAMGHZM one can interpret the HVs from the QAD are "fixed" with HVs that emerge from the DD, just like the HVs in the latter can "fix" its three quanta, it not surprising that the device operates as expected. Consequently, the added QAD allows to explain the MGHZM, in terms of its own HVs, without contradiction, when the four types of SCs are imposed.

[1] This is a slightly modified version from the original.

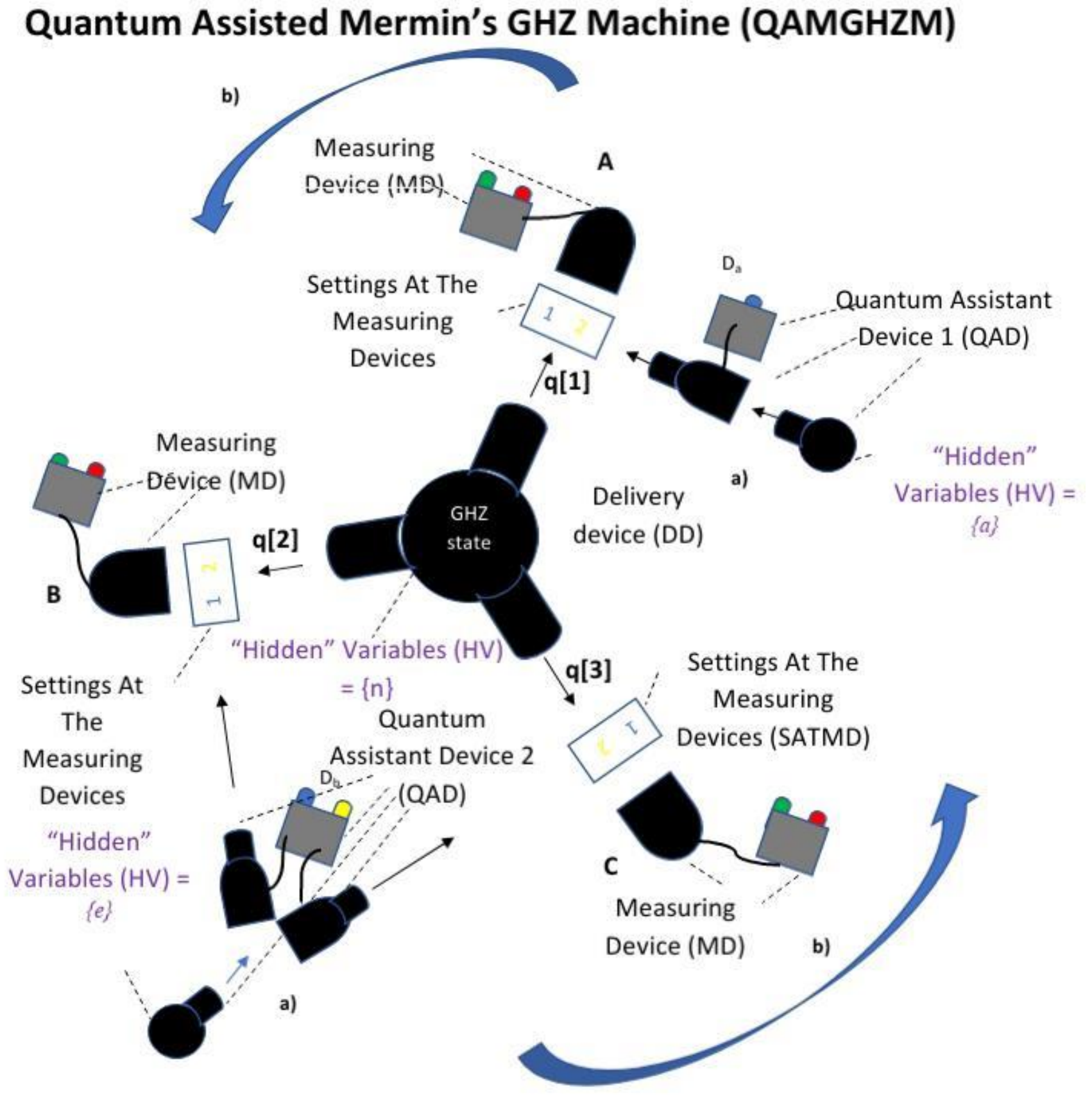

**Figure 2**. **a.** QAD 1 determines the setting for the quantum that arrives at detector A; $D_a$ always flashes blue, activating setting 1; QAD 2 determines the settings for B and C; when $D_B$ flashes blue, setting 1 is imposed on B and C; a yellow flash keeps the default setting at 2 on both; Settings cannot be observed directly, only implied by the output of $D_A$ and $D_B$. **b.** Both assistants are rotated, after collecting data, for two more experiments where detector B and C also have constant setting 1, while C,A and A,B could have setting 1 or 2 with QAD 2.

### B. A restricted version of quantum superdeterminism

The above analysis of the quantum assisted situation can be characterized as **quantum superdeterminism** (QSD) given that there is no assumption that the GHZ state information, before encountering the settings, is independent of the type of settings that were going to be implemented [2-3]. In general, one may always interpret that the "random choice" of settings (like those the scientists make in the MGHZM), somehow, are correlated with the HVs in the GHZ state so that the expected results are detected every time measurements are performed in seemingly "arbitrary" ways. In this perspective, when the scientists make a SC selection, somehow the GHZ state was already coordinated with their "choices", even before they implemented their "decisions", as if the scientists, the devices, or something else in the environment, were a quantum system that became somehow correlated with the GHZ state. On the other hand, it is often unclear how to relate a non-quantum system assisting with SCs on the GHZ state, as if both were an entangled-like state, because as far as the scientists doing the experiment know, they are not "conspiring" with the rules of the MGHZM, don't have any knowledge about the HVs, and the SATMD devices, or any other "classical system" (i.e., neglecting miniscule quantum mechanical effects on those "classical systems"), are acting independently of the GHZ state. To get around the issue, the QAD allows to include its HVs with those from the GHZ state, to substitute the scientists, or any "classical system" making "random" selection of SCs. More explicitly, the QAD

(i) makes all the "choices" of SCs that lead to different outcomes,

(ii) Imposes the SCs directly,

(iii) informs the type of SCs that were implemented,

which allow to focus into two quantum systems (the quantum-assistants) that are under experimental control. In contrast, the scientists, the SATMD, and other aspects of the environment that may have led to a particular SC in the MGHZM cannot be considered explicit variables in the QAMGHZM. The QAMGHZM allows to isolate the HVs that are already correlated with those from the GHZ state, and anticipate the information, instead of trying to apply the QSD interpretation to the selection process and implementation in the MGHZM.

In this manner, the QAMGHZM limits the scope of QSD and avoids the eccentric aspects that can be argued to be unscientific[2]. One common criticism of QSD is that,

*"In any scientific experiment in which two or more variables are supposed to be randomly selected, one can always conjecture that some factor in the overlap of the backward light cones has controlled the presumably random choices. But, we maintain, skepticism of this sort will essentially dismiss all results of scientific experimentation. Unless we proceed under the assumption that [superdeterminism does not hold,] we have abandoned in advance the whole enterprise of discovering the laws of nature by experimentation."* [4].

Any "conspiracy" theory can be created when there is nothing empirical suggesting that it must be so. Suggesting, WITHOUT evidence, that any "random" mechanism making the "choices" of SCs on the GHZ state is correlated already with the HVs, in the MGHZM example, is as unscientific as suggesting, WITHOUT evidence, that voting machines are arranged to favor a particular candidate somehow. There is no compelling reason, one might say, to suggest that the HVs in the GHZ state, in the MGHZM analysis, are somehow arranged with any process of SC selection, carried out by, let's say, a scientist, so that both are really not independent events. In contrast, given that the results were correlated among all the quanta of the QAMGHZM (for the four types of settings), and there is no way to distinguish anything special among them from an instrumentalist perspective because their outcomes were the only way to verify the machine's performance, there is evidence implying the likelihood that the interdependence of the information in the QADs with the GHZ state, could be of the same physical nature as the GHZ state alone in the MGHZM (entanglement-like), in every instance there is a joint measurement. In other words, it is not surprising that all the quanta seem to have entanglement-like correlations with one another including the quantum-assistants. Moreover, one could interpret the "rules" of the QAMGHZM as if they were established before any run of the experiment took place, without logical contradiction, which is something that cannot be affirmed in the MGHZM setup, where the assumption of "hidden" rules itself lead to a contradiction, and it is not always clear how the process of setting selection would be arranged with the GHZ state, as implied by QSD. For instance, if three scientists agreed to change settings suddenly, in the MGHZM situation, what exactly would be correlated with the GHZ state so that there is statistical interdependence? The scientist? The MD? The SATMD? All of them? Something else in the environment? When there is a QAD, the answer is simply the HVs of the quanta assisting with the SC. The quantum assisted setup allows to resolve the "mystery" of the MGHZM, in terms of HVs, by shifting the "mystery" to the way the QAD's information could have been arranged with the GHZ state itself before the experiment began. After all, it is well known that quanta can have correlations with other quanta [5] in ways that classical systems do not. Assuming that the information in the QAD has such a correlation with the entanglement is not eccentric at all, and simply an extension of the idea of entanglement. It is not unscientific to induce that other quanta may have entanglement-like properties with other space-like separated quanta even if they do not seem to have interacted with each other; certainly, that is one possible interpretation of quantum swapping experiments [6-7].

[2] See acknowledgement

In contrast, in situations where the setting selection process is a function of some classical mechanical system, that delivers and informs the settings, there is no need to introduce interdependence among the HVs of the GHZ state with that classical system because the latter does not have "hidden" information like the quantum-assistants. Of course, there can be semi-classical situations where quantum effects become significant in a relatively small time even though the situation involves macroscopic objects, as is the case of billiard balls on a pool table [8], or chaotic systems [9], where the significance of those quantum effects has to be assessed empirically. However, those semiclassical situations might satisfy requirement (i), if they are somehow used as a selection mechanisms, but it is not clear how the other requirements might be satisfied given that there is the high probility that the quantum information might be "destroyed" before it can inform. Now, experiments have been realized where quanta that arrive from distant, separate sources, end up making "the choices" of settings, but, in those, it is a classical system that applies the settings [10], which necessarily has the effect of "deleting" the information from the quanta assisting with the selections, which necessarily implies that (ii) and (iii) are not satisfied. The same can be said about any conjecture that might imply that a quantum mechanical process in the brain of the scientist [11] could lead to specific selection of settings. There are quantum correlations among HVs of superposed states, from different quanta, even if they are space-like separated, but there is no such a correlation between a quantum and a classical system in the restricted version of QSD introduced. In this way, by restricting QSD to the situation involving quantum-assistants only, that satisfy (i), (ii), and (iii), it is still possible to perform experiments that maintain the independence of the experiment from the results, which is necessary to "advance the whole enterprise of discovering the laws of nature by experimentation". In this way, the QAMGHZM allows to interpret an entanglement-like relation between the quanta involved in the selection process and the GHZ state, that was there already before measurements, while assuming that any classical mechanical situation can function independently of another, such as in assemblying the equipment to carry out an MGHZM-type of experiment to find out the quantum sub-systems that may have participated in the selection process (if that were the case) that can be regarded to be quantum-assistants.

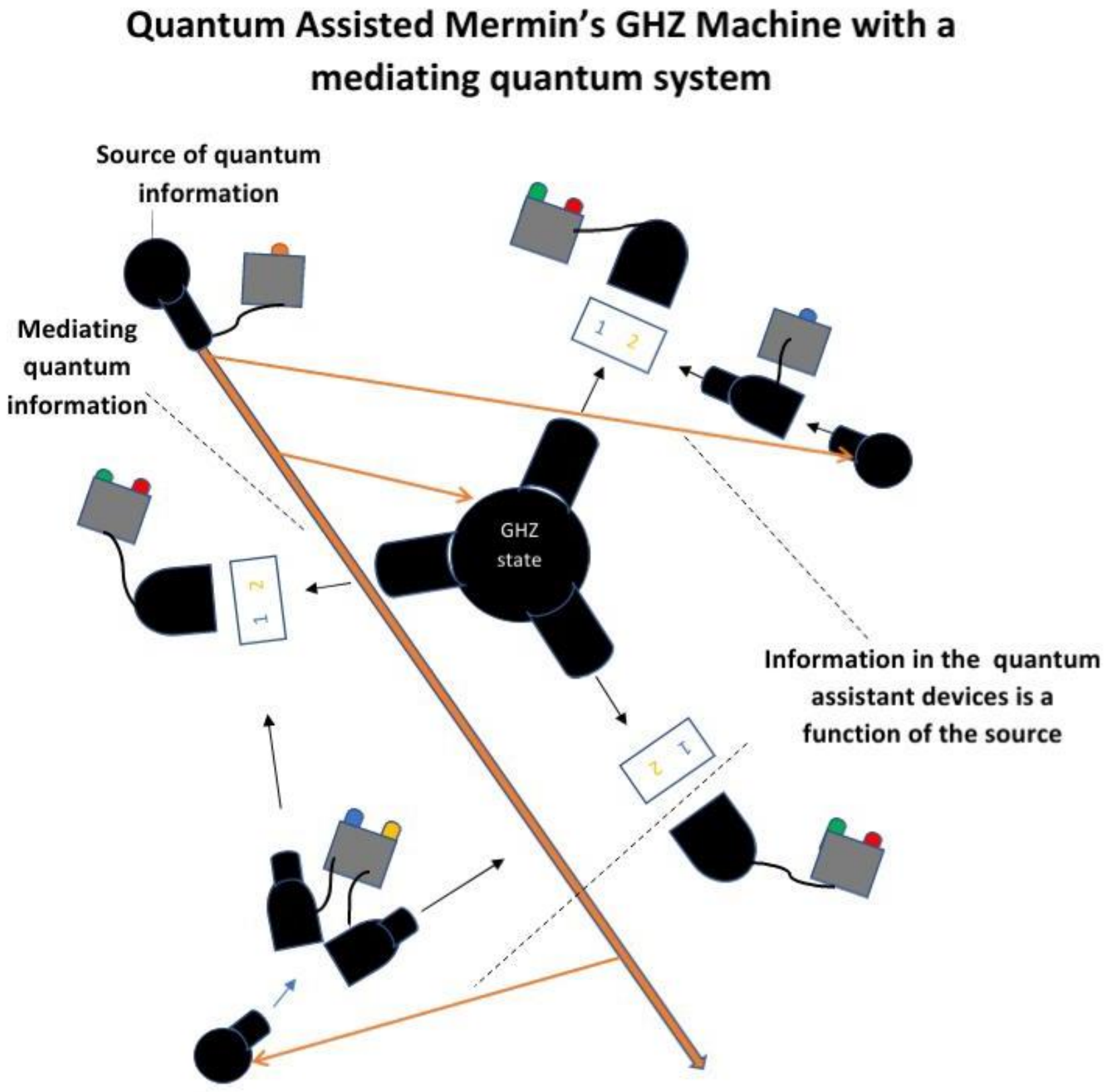

**Figure 3.** The information from the DD and the quantum-assistants could be functions of a common quantum system. Every orange flash is associated with each joint measurement. The mediating system can be regarded to originate the correlation between the sets of "hidden" variables. In this way, the "hidden" variables could be interpreted as being local, and through experimentation such a hypothesized common quantum system could be verified or ruled out.

## C. The hidden variables may be local

Although the entanglement-like relation between the information in the QAD and the GHZ state can be argued to be merely a mathematical convenience, it is also possible that the correlation may have been originated by a mediating, local system, which would entail that their HVs are also local as the result of a previous physical process that originated the arrangement. One may object that the QAD is not connected with the DD, and thus the "rules" originated at the DD are exclusive. One may ask, how would the quanta from the QAD be initialized in coordination with the quanta from the DD, if they were not constructed together as is the case with the information prepared within in the DD? Is there something passing information among them? One may respond that the correlation of all the quanta involved was established BEFORE the experiment took place, through events in space-time just like the DD organized the HVs for only its three quanta; such a hypothesis could be tested by identifying any quantum system that may have had a common correlation with the HVs of the entanglement and the information in the QAD before every joint measurement. Along this line of inquiry, David Bohm [12] seemed to have considered that a mediating system, between two other systems that can be considered "classical", can be expected whenever there is a "link" between the two, which would contrast quantum systems when they are not considered to be "separate" parts that can be assigned something like HVs that have interdependence; however, because in the QAMGHZM, the quantum systems can be considered to have HVs, as if they were "separate" parts relative to the others, a mediating system can be expected if those HVs and their "link" (the correlation) are local and not just a mathematical convenience. In this way, if no mediating quantum system could be associated with both the GHZ state and the information in the QAD before making a joint measurement then one could say that assuming a correlation between them before making those measurements is merely a mathematical tool. However, any common quantum system that had a physical relation with both the information in the DD and the QAD can be interpreted to have provided the arrangement. As shown in fig.3, a new setup for the QAMGHZM can be made where an additional device gives orange flashes every time an additional quantum is delivered, which ultimately had a functional relation with at least one quantum from each device without "destroying" their information. Without identifying the additional mediating quanta, the scientists running the experiment might be puzzled by the QAMGHZM setup where it seems that the devices work in coordination but without knowing how it was achieved. Such a mediating system could be interpreted as the carrier of the "rules" that were transferred to all the devices and the correlation is the effect; thus, the statistical interdependence of all the devices may be a local physical event. In this manner, it is possible that before running each experiment, the GHZ state and the assistant systems were functions of some quanta that can be identified through experimentation; however, such a hypothesis is falsifiable.

Moreover, one may think that all of the experiments run with the QAMGHZM disclose the HVs, of the GHZ state, in relation to those from the QAD, whereas the MGHZM makes them fuzzy from the point of view of the scientists, and surrounding classical systems, that may have participated in the changes made to the SCs. The HVs of the QAD provided a reference frame (in an analogy with general relativity involving HVs [13]) from which two different points of view can result out of the two ways to transform the coordinates (i.e., "hidden" rules); any common quantum system that could have been associated with both, before every joint measurement, can be said to have provided the relation between their coordinates (like a mass provides the "curvature" of space for the transformation between two reference frames in its surroundings). Thus, it is reasonable to assume that the entanglement-like correlation among the HVs is physically meaningful if one can find a common quantum system with which both can be said to have been correlated and that those HVs can only exist in relation to others.

## D. The three quantum computer experiments

The quantum computer equivalent of the QAMGHZM is presented in fig. 4; the three experiments are quantum assisted versions of other GHZ-like experiments that have been already realized in a superconducting quantum computer [14]. The experiments were realized on an IBM quantum computer with 5-qubits, where measurements were done sequentially in such a way that it was possible that the phase-gate for one of the qubits in the GHZ state was set before the phase-gates for the other two entangled qubits, that paralleled the situation depicted in “Quantum mysteries revisited”. In this way, two entanglements were obtained in each experiment, where one of them was a linear combination of four tensor products, in the z-basis (“up” state=|0>, “down” state=|1>), containing an **Odd Number Of “Down” States** (ONODS), in each tensor product term, let’s call it an **Odd Entanglement** (OE), and another containing a **Non-Odd Number Of “Down” States** (N-ONODS), which will be referred as a **Non-Odd Entanglement** (N-OE). It is important to mention that q[0] and q[4] were not entangled to the rest of the qubits, but only served to select the type of settings. In each of the experiments, the same phase-gate ($\boldsymbol{P}(-\frac{\pi}{2})$) under the control of q[0], was set at least one time, on different qubits, and q[4] set, or not, two phase-gates on the other entangled qubits, mirroring the situation presented in “Quantum mysteries revisited” where, in three of the possible settings, only one operator was different from the other two operators, and in only one, the same type of operator was used for all three qubits. The set of four possible SCs can be expressed as

$$\{\boldsymbol{P}_{q[1]}\left(-\frac{\pi}{2}\right)\boldsymbol{I}_{q[2]}\,\boldsymbol{I}_{q[3]}\,;\,\boldsymbol{I}_{q[1]}\,\boldsymbol{P}_{q[2]}\left(-\frac{\pi}{2}\right)\boldsymbol{I}_{q[3]}\,;\,\boldsymbol{I}_{q[1]}\,\boldsymbol{I}_{q[2]}\,\boldsymbol{P}_{q[3]}(-\frac{\pi}{2})\,;\boldsymbol{P}_{q[1]}\left(-\frac{\pi}{2}\right)\boldsymbol{P}_{q[2]}\left(-\frac{\pi}{2}\right)\boldsymbol{P}_{q[3]}(-\frac{\pi}{2})\},$$

where $\boldsymbol{I}$ is the absence of a $\boldsymbol{P}(-\pi/2)$ gate on the corresponding qubit. The results from those SCs were two entanglements, where measurements of q[0] and q[4] allowed to distinguish whether the SC belong to the subset

$$\{\,\boldsymbol{P}_{q[1]}\left(-\frac{\pi}{2}\right)\boldsymbol{I}_{q[2]}\,\boldsymbol{I}_{q[3]}\,;\,\boldsymbol{I}_{q[1]}\,\boldsymbol{P}_{q[2]}\left(-\frac{\pi}{2}\right)\boldsymbol{I}_{q[3]}\,;\,\boldsymbol{I}_{q[1]}\,\boldsymbol{I}_{q[2]}\,\boldsymbol{P}_{q[3]}(-\frac{\pi}{2})\},$$

where each element will be referred as a **Non-Odd Setting Combination** (N-OSC), or the subset

$$\{\,\boldsymbol{P}_{q[1]}\left(-\frac{\pi}{2}\right)\boldsymbol{P}_{q[2]}\left(-\frac{\pi}{2}\right)\boldsymbol{P}_{q[3]}(-\frac{\pi}{2})\},$$

the only **Odd Setting Combination** (OSC); however, three experiments had to be performed to obtain results with all four possible settings (three setting combinations had only one phase-gate, located in each of the three qubits once, while the other setting combinations had all three). Each experiment resulted in either OEs or N-OEs, with equal frequency, such that N-ONODS resulted from all entangled qubits having only one phase-gate, and ONODS resulted from all qubits having just one. Because three out of the four possible settings corresponded to one phase-gate set on one of the three entangled qubits, and only one setting corresponded to all qubits having one, the results from all three experiments gave an excess data for the only OSC.

## Circuits for Quantum-assisted Mermin's GHZ Machine

### Experiment 1

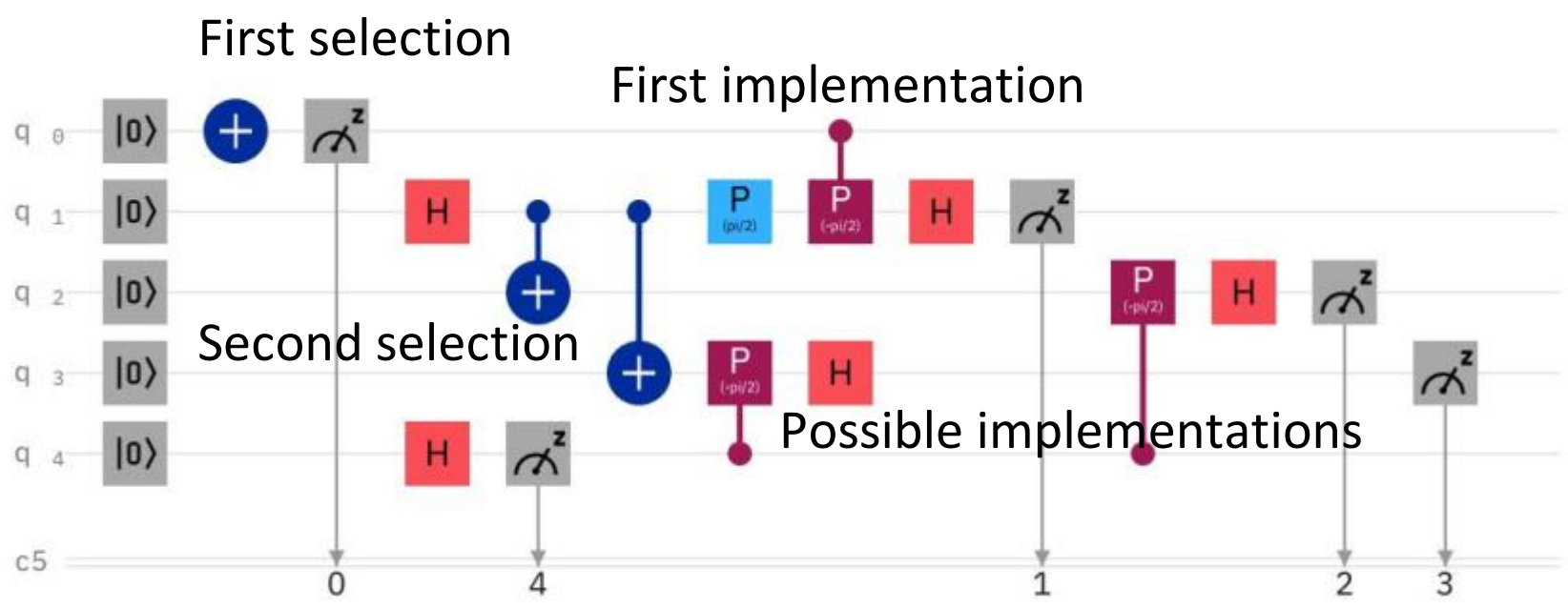

### Experiment 2

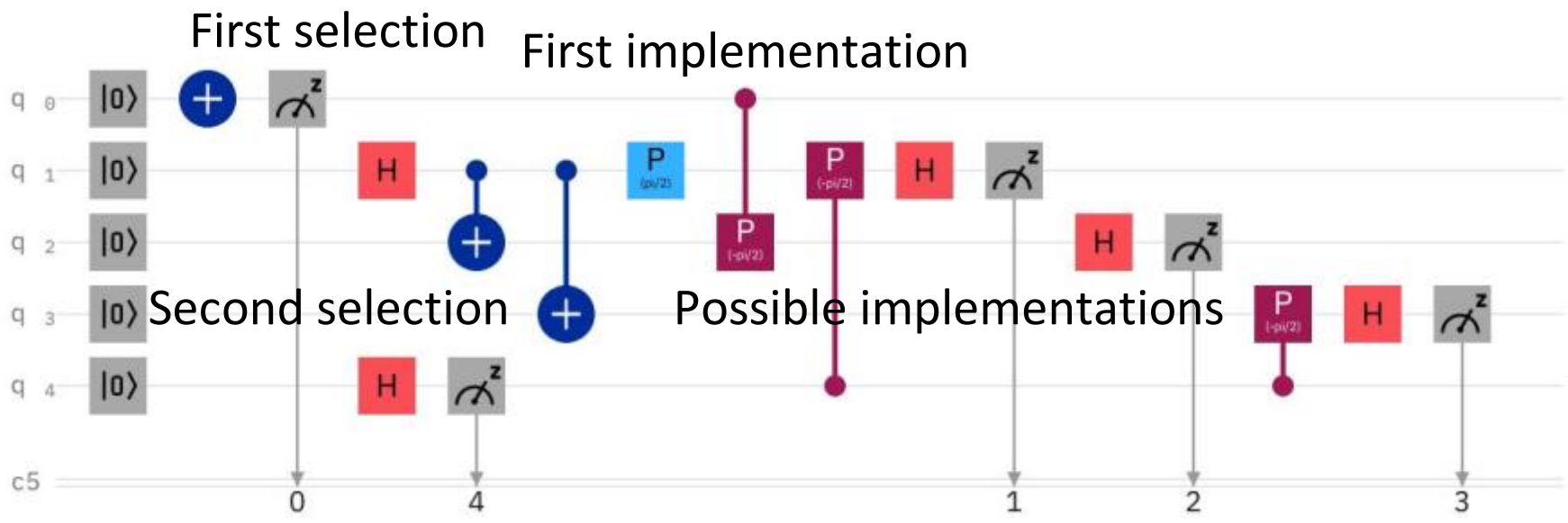

### Experiment 3

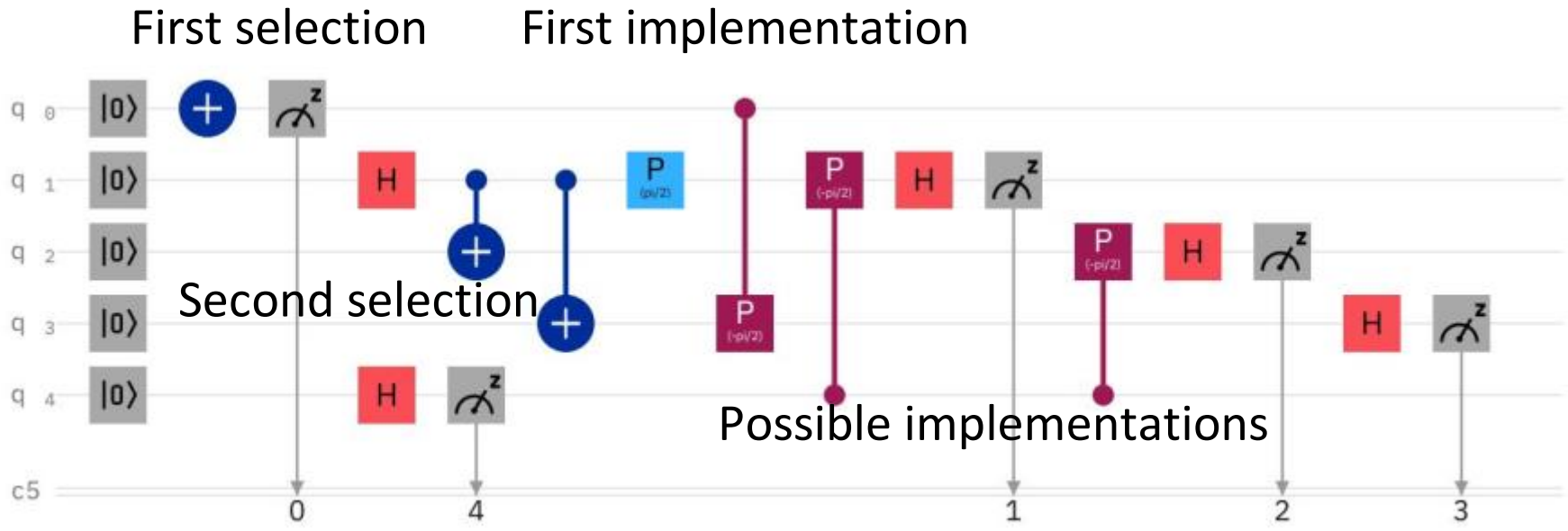

**Figure 4.** First, q[0] was measured, resulting always in a phase-gate being imposed on either q[1], q[2], or q[3], corresponding respectively to experiment 1,2, or 3. Then, a second measurement, imposed, or not, two phase-gates on the other two qubits that formed the GHZ-like state. The outcomes from each experiment could be either a non-odd number of "down" states, or an odd number of them. In this way, the selection of the type of setting combinations, that led to a specific type of outcomes, is done by the assistant qubits q[0] and q[4] in each experiment.

The assisted setup in fig. 4 will be referred as a DCOS experiment given that there was a time difference, with unknown uncertainty, between the selection of the first gate and the selection of the other two, without knowing, from the start of the experiment, if those two were going to be selected during any "shot". If the time when the phase-gate is selected to be on one of the entangled qubits is given by $t_i^{(selection)}$, then the time when q[4] selected the two phase-gates (and obtain an OE) is

$$t_i^{(selection)} + \Delta t_i^{(selection)} , \tag{1}$$

where $\Delta t_i^{(selection)}$ is the time difference between the selection of the first gate and the other two, in every OE measurement, for the $i^{th}$ experiment (a total of three). $\Delta t_i^{(selection)}$ was the same in each of the three experiments; they were estimated by the number of gates between the selection of the first and the other two in the transpiled circuit (fig. 11 in Sec II. D). Also, the transpiled circuit shows the order in which those phase-gates were implemented: the same order in which the selections were made. In this way, the settings were "unpredictable" for any one type of HVs in the entanglement, without having correlation to the HVs of the assistant qubits, to be able to have simulated the results, as was the case in the analysis of the MGHZM. In this way, the experiments can be considered DCOS because there was a process delay, with unknown uncertainty, between the specific selections that occurred after the experiment had started, that increased the likelihood that those selections were not done as close in time as possible.

Also, any "classical" interpretation could be challenged, even if the selections had occurred simultaneously in some measurements, given the qubits were space-like separated and, apparently, without all of the relevant qubits exchanging information neither during the selection process nor the implementation; as shown in fig. 4 (and fig. 11), there were no multiple-qubit operators, AFTER the experiment had began, that could be interpreted as a path for communication between relevant qubits. The fact that the entangled qubits, in any run of the experiment with OE, did not have the SCs completely implemented from the start, mirrors the situation depicted in the analysis of MGHZM where SCs can change in "mid-flight", knowing that they became somehow implemented before measurements of the entanglement. In this way, rather than having the settings being established from the start, as it is often done in MGHZM-like experiments performed on superconducting quantum computers [15-18], the SCs in the present paper mirror Aspect-like experiments [19], but with three entangled qubits with just two SCs in each experiment (a total of four SCs for all three experiments), where the settings were "randomly", quickly varied while each was running, without the scientist knowing, from the beginning, the type that were going to be selected, except that the purpose for the variations in the present experiments was not an attempt to rule out superluminal communication among the qubits; the goal was to substitute the scientist from the selection process that leads to a specific type of outcome with a quantum system (either N-OE or OE), where such a selection occurs once the experiment had already started and the quantum-assistants also inform the type of settings that were applied.

### E. The assumption in Mermin's inequality.

Now, to compare the results obtained with the circuit in fig. 4 with a typical non-quantum assisted version of Mermin's GHZ-like experiment, **Mermin's Inequality** (*MI*) (for the case of three quanta) was used to verify the expected "violation", by first assuming that the SCs were imposed independently of the entanglement, a necessary assumption to derive MI; then, those assumptions were evaluated with the QSD interpretation to show that, in the quantum assisted case, *MI* does not follow, and the HV assumption is consistent with the quantum theory in these particular experiments. The expression for Mermin's polynomial, for three quanta 1,2,3, and given two operators a and a' [20-21], is

$$\boldsymbol{M_3 = a_1' \cdot a_2 \cdot a_3 + a_1 \cdot a_2' \cdot a_3 + a_1 \cdot a_2 \cdot a_3' - a_1' \cdot a_2' \cdot a_3'}. \tag{2}$$

*MI* is the expression

$$< \boldsymbol{a_1' \cdot a_2 \cdot a_3} > + < \boldsymbol{a_1 \cdot a_2' \cdot a_3} > + < \boldsymbol{a_1 \cdot a_2 \cdot a_3'} > - < \boldsymbol{a_1' \cdot a_2' \cdot a_3'} > \leq \boldsymbol{2} \tag{3}$$

which follows from "classical" considerations like any other bell-type inequality. In the present experiment the first three terms of (3) correspond to only one phase-gate set on one of the qubits, while the last term corresponds to all three phase-gates set on all of the qubits. By treating the output obtained from the quantum computer in terms of "0's" and "1's" to be "$+1$'s" and "$-1$'s", each term in (3) expresses the average "product" that results from each of the four types of settings. Each of the first three, from a quantum perspective, should result in "$+1$"; the last term should be "$-1$"; thus, the quantum theory predicts $+4$ when all averages are combined, which is greater than the value predicted by the inequality under "classical" assumptions. Using the circuits in Fig. 4, the value obtained for the left-hand side of (3) was 2.235863182 which allows comparison with other results obtained by non-quantum assisted means. Now, those assumptions needed to "derive" the inequality are the same as those used in the MGHZM analysis: the process for implementing SCs is not correlated with the HVs in the entanglement. Certainly, results from the present experiment could be interpreted to "violate" *MI* when measurements of the entangled qubits were performed in four different ways (as implied in the expression (3)), with the assistance of q[0] and q[4] rather than with the assistance of an "external" agent that could select and implement the SCs, under the assumption that there is statistical independence between the outcomes for the settings and those for the entanglement. The independence of the process for SC selection and the HVs of the entanglement is a typical assumption implied in the published experiments, using superconducting quantum computers [15-18], that simulate MGHZM-type of experiments, where each separate setting was set by the subject running those experiments (not through quantum assistance as defined in Sec. IB). In those non-quantum-assisted cases, the "decisions" were done by an "outside" agent that cannot be accounted in the analysis. Nevertheless, the present experiment, could also be analyzed in terms of inter-correlated HVs, even if there is "violation" of MI, because those for q[0] and q[4] can be interpreted to anticipate the future outcome of q[1],q[2] and q[3]and adjust to the type of entanglement, before any measurements were performed, dismissing the assumptions necessary to "derive" MI (that there is independence between the process for SC selection and the HVs in the GHZ state). That is, under the QSD interpretation, if the "hidden" variables for q[0] and q[4] are *{a}*,*{e}* respectively, and those for the entanglement are *{n}*, one can expect the partitions $\{e^{(+)}\},\{e^{(-)}\}\ of\ \{e\}$, and the partitions $\{n^{(+)}\},\{n^{(-)}\}\ of\ \{n\}$, to be correlated in two different ways:

$$\{\boldsymbol{n}^{(+)}\}\boldsymbol{R}\{\boldsymbol{e}^{(+)}\} \Leftrightarrow \text{Non-odd number of } |1> \qquad (4)$$

$$\{\boldsymbol{n}^{(-)}\}\boldsymbol{R}\{\boldsymbol{e}^{(-)}\} \Leftrightarrow \text{odd number of } |1> \qquad (5)$$

such that $\boldsymbol{R}$ relates the type of HVs in each "shot", before SCs are imposed, where $\{n^{(+)}\}\boldsymbol{R}\{e^{(+)}\} \equiv \{e^{(+)}\} \times \{n^{(+)}\}\ and\ \{n^{(-)}\}\boldsymbol{R}\{e^{(-)}\} \equiv \{e^{(-)}\} \times \{n^{(-)}\}$, which necessarily results in the quantum theory predictions. Expressions (4) and (5) will be referred as *arranged (or correlated) relations*. Notice that changes of HVs from the set $\{a\}$ do not have an effect on the output, i.e., q[0] imposes the same phase-gate in all three experiments, but on a different qubits in each experiment. In contrast, the output of q[4] could result in OEs or N-OE. The relations are illustrated in fig. 5. Now, without $\{n^{(+)}\}\boldsymbol{R}\{e^{(+)}\}\ and\ \{n^{(-)}\}\boldsymbol{R}\{e^{(-)}\}$, the sets $\{n^{(+)}\},\{n^{(-)}\}$ result in non-odd or odd entanglements without having a clear relation to the way the settings were imposed, as is the case the non-assisted experiments. One can say that $\{n^{(+)}\}\ and\ \{n^{(-)}\}$, in the non-assisted case, coincide with the "correct" theoretical settings, WITHOUT being able to state a correlation before measurements (fig. 6). Also, it is possible to regard no special relation among the HVs in the quantum assisted version with the quantum computer, i.e., the relation between the HVs is simply $\{n\}\boldsymbol{R}\{e\} \equiv \{e\}x\{n\}$ which merely entails that the results must satisfy MI; the relations are illustrated in fig. 7 and will be referred as a *disarranged relation*. Nevertheless, the results obtained from the three experiments show violation of MI, which entails that $\{n^{(+)}\}\boldsymbol{R}\{e^{(+)}\}$ *and* $\{n^{(-)}\}\boldsymbol{R}\{e^{(-)}\}$ explain the results, rather than $\{n\}\boldsymbol{R}\{e\}$, from the start of each "shot".

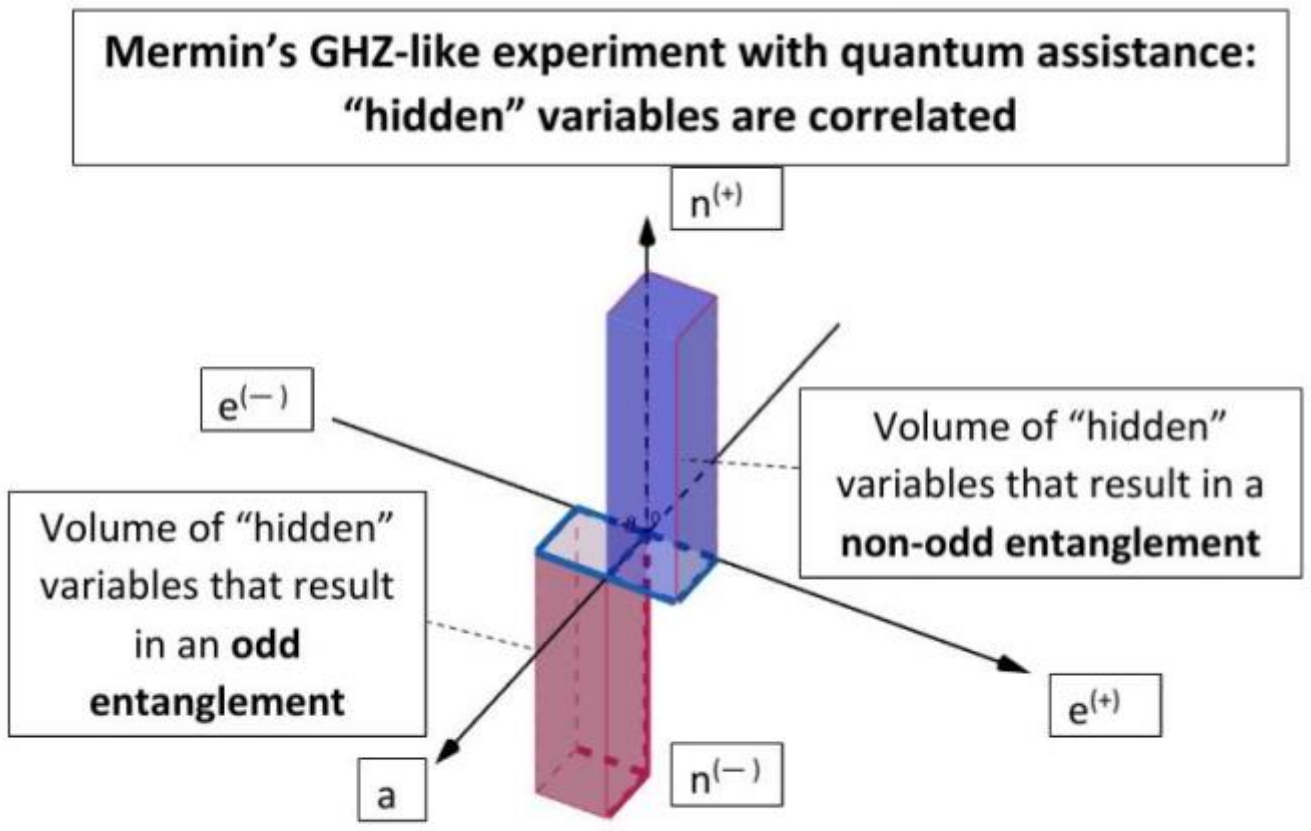

**Figure 5.** In the quantum assisted version of Mermin's GHZ-like experiment, one can assume that the "hidden" variables of the entangled quanta and those for the assistant system, are correlated in such a way that they belong to two specific volumes of the "hidden" variable space as shown. In other words, when the settings result in non-odd entanglements, the quantum-assistants necessarily result in the corresponding setting combination on all three qubits before any measurements as if the GHZ state had information about the settings that were going to be imposed; same when an odd entanglement is the outcome.

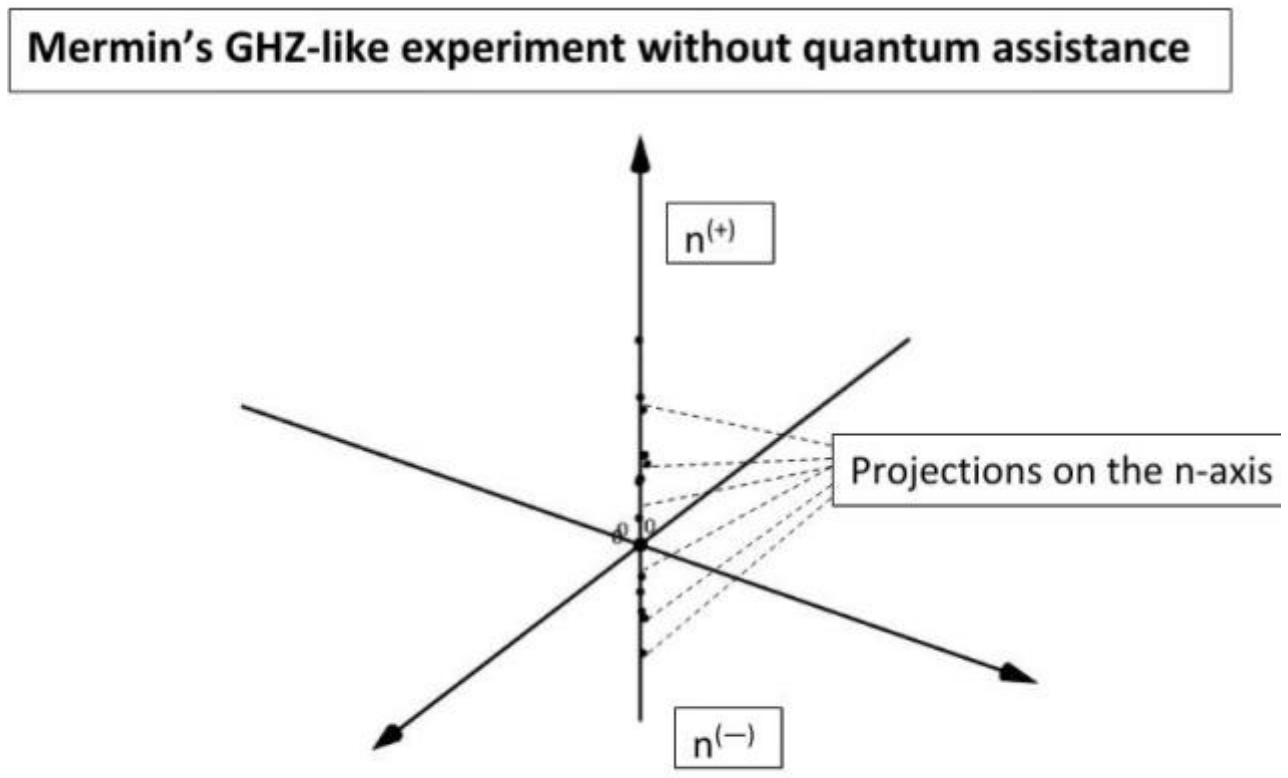

**Figure 6.** In a typical experiment with a GHZ-like state where four settings are "randomly" selected to calculate Mermin's inequality (for three quanta), the process leading to those settings is not explicitly stated from a quantum mechanical perspective. Without such a "reference frame", which is an extra dimension in the "hidden" variable space, only the projection on the entanglement "hidden" variable's axis can be implied from the measurements (either a non-odd or odd number of "down" states), even though the MGHZM device works.

The above analysis of the quantum assisted experiment in terms of Mermin's inequalities can be made explicit by writing each of the qubit functions in terms of the HVs. In the quantum assisted case, the set of functions for those "hidden" variables is

*(6)*

$$\left\{\boldsymbol{S'}_{q[\beta']}(a,t), \boldsymbol{F}_{q[b]}\left(n, \boldsymbol{S'}_{q[b]}(a,t)\right), F_{q[c]}\left(n, \boldsymbol{S}_{q[c]}(a,t)\right), F_{q[d]}\left(n, \boldsymbol{S}_{q[d]}(a,t)\right), \boldsymbol{S}_{q[\beta]}(e,t)\right\},$$

$$such\ that\ b,c,d \in \{1,2,3\}, b \neq c \neq d\ , \qquad c = b+1\ mod\ 3, \qquad d = b+2\ mod\ 3;$$

the functions $\boldsymbol{F}_{q[1]}, \boldsymbol{F}_{q[2]}, \boldsymbol{F}_{q[3]}$ operate on q[1], q[2], and q[3] respectively, and $\boldsymbol{S'}_{q[\beta']}, \boldsymbol{S}_{q[\beta]}$ are the functions of the HVs of q[0] and q[4] acting on different qubits (either q[0], q[1], or q[2]). Notice that the functions for the entanglement can be written in two ways with the partitions of *n* alone, which necessarily result in an ONODS and N-ONODS respectively:

*(7)*

$$\left\{\boldsymbol{F}_{q[b]}\left(n_i^{(+)}, s_{q[b]}\right), \boldsymbol{F}_{q[c]}\left(n_i^{(+)}, s_{q[c]}\right), \boldsymbol{F}_{q[d]}\left(n_i^{(+)}, s_{q[d]}\right)\right\}$$

*or*

$$\left\{\boldsymbol{F}_{q[b]}\left(n_j^{(-)}, s_{q[b]}\right), \boldsymbol{F}_{q[c]}\left(n_j^{(-)}, s_{q[c]}\right), \boldsymbol{F}_{q[d]}\left(n_j^{(-)}, s_{q[d]}\right)\right\},$$

such that b, c, d ∈ {1,2,3}, b≠c≠d, c≡b+1 mod 3, d≡b+2 mod 3, and $s_{q[b]}$, $s_{q[c]}$, $s_{q[d]}$ are the actual settings imposed. Three of the settings for Mermin's polynomial correspond to the first set (the N-OSCs); only one corresponds to the second set (the OSC); by regarding the "up" and "down" states to be "+1" and "-1" respectively, one can easily demonstrate that the average of Mermin's polynomial is greater than 2, which is necessarily a "violation" of MI, but it is consistent with the quantum theory.

Now, using the language of the EPR paper [5], one may say that just by considering the entanglement's HVs, in the MGHZM analysis, is NOT a complete description of "reality"; however, this is not the case in the assisted-quantum experiment, because the outcomes of those assistant qubits and the entanglement can be interpreted consistently in terms of HVs from the start, without qubit communication while running the experiment. From this perspective, one may interpret each of the computational basis output elements from the quantum computer as having been arranged before any of the measurements were performed. Qubits could be interpreted to have information of the final state of all the other qubits even before all quantum gates had processed all the information; violation of Mermin's inequality does not imply that there are no "hidden variables" in the quantum assisted experiments. In this way, the initial conditions are defined in such a way that, no matter what the settings were going to be before each "shot", the HVs in the qubits that define those settings, and those in the entangled qubits, were arranged to provide self-consistent information no matter when the SC were imposed or if they were imposed by delayed-choice, as have been suggested with comprehensive models [22-23], or in a way that can be implemented in others that simulate, deterministically, some aspects of quantum systems "classically" [24] that require strategic initial conditions to deal with the "spooky" aspects that the present paper resolves in terms of pre-arranged relations among the "hidden initial conditions" that may also be local if those relations result from previously local arrangements.

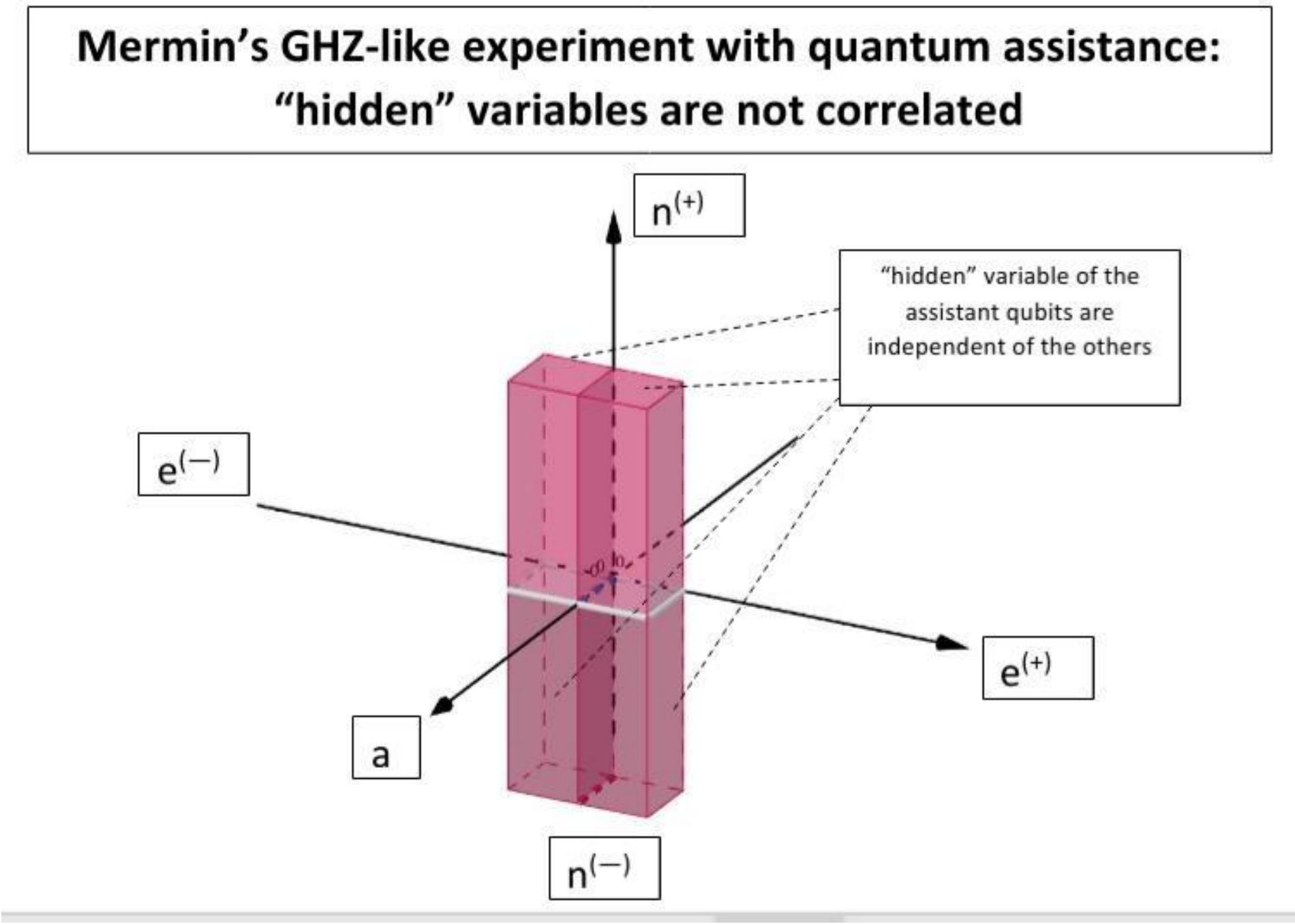

**Figure 7.** The quantum-assisted version of the MGHZM-like experiment with the quantum computer can also be interpreted as if the "hidden" variables from the assistant system and the entanglement are statistically independent of one another. In this perspective, Mermin's inequality should not be "violated": the assumption necessary to "derive" the inequality is that any of the four relations depicted in the four rectangular prisms can occur (i.e., setting selection is "random" and not conditional upon the "hidden" variables of the entanglement); however, computation of the inequality with the entanglement measurements for the four settings surpass the limit set by the inequality.

# II. Method

## A. Analysis of the quantum circuits.

To obtain robust and meaningful data, 8000 shots of the experiment were performed in each run with the three circuits shown in fig. 4. There were two settings in each experiment: one for just one phase-gate on one of the qubits forming the entanglement, and another for phase-gates on all of the entangled qubits; the essential difference between the three experiments is the qubit where the one phase-gate was placed. Mermin's polynomial was calculated by regarding the measurements obtained from the entanglement as if they were "up" or "down" spin-1/2 states i.e the states |0>, |1> corresponded to +1, —1 respectively. That is, for each of the four possible SCs (Each experiment contained two SCs), the average product of their corresponding "spins" was calculated. Because there was only one OSC, but there were three types N-OSCs, the number of "shots" with OSCs was, in principle, three time the number for each type of N-OSCs after all three experiments had been performed.

As shown in fig. 4, q[0] and q[4] were used as the selection qubits, and q[1], q[2], and q[3] were in the GHZ-like state

$$|GHZ >= \frac{1}{\sqrt{2}}(|000 > +i\ |111 >);$$

the outcome of the selection process resulted in either only one **P (**$-\frac{\pi}{2}$) gate imposed on the entanglement, or three P **(**$-\frac{\pi}{2}$) gates; then, **H** gates were applied to the three qubits before they were measured. Formally, the operations involved in the selection process can be expressed as matrices

$$\boldsymbol{H} = \frac{1}{\sqrt{2}}\begin{pmatrix} 1 & 1 \\ 1 & -1 \end{pmatrix},$$

$$\mathbf{P}\ (-\frac{\pi}{2}) = \begin{pmatrix} 1 & 0 \\ 0 & e^{-i\frac{\pi}{2}} \end{pmatrix},$$

$$\text{c}\mathbf{P}\ (-\frac{\pi}{2}) = \begin{pmatrix} 1 & 0 & 0 & 0 \\ 0 & 1 & 0 & 0 \\ 0 & 0 & 1 & 0 \\ 0 & 0 & 0 & e^{-i\frac{\pi}{2}} \end{pmatrix},$$

such that

$$|0> = \begin{pmatrix} 1 \\ 0 \end{pmatrix}, |1> = \begin{pmatrix} 0 \\ 1 \end{pmatrix},$$

and

$$\begin{pmatrix} 1 \\ 0 \end{pmatrix}\begin{pmatrix} 1 \\ 0 \end{pmatrix} = \begin{pmatrix} 1 \\ 0 \\ 0 \\ 0 \end{pmatrix}, \begin{pmatrix} 1 \\ 0 \end{pmatrix}\begin{pmatrix} 0 \\ 1 \end{pmatrix} = \begin{pmatrix} 0 \\ 1 \\ 0 \\ 0 \end{pmatrix}, \begin{pmatrix} 0 \\ 1 \end{pmatrix}\begin{pmatrix} 1 \\ 0 \end{pmatrix} = \begin{pmatrix} 0 \\ 0 \\ 1 \\ 0 \end{pmatrix}, \begin{pmatrix} 0 \\ 1 \end{pmatrix}\begin{pmatrix} 0 \\ 1 \end{pmatrix} = \begin{pmatrix} 0 \\ 0 \\ 0 \\ 1 \end{pmatrix}.$$

It is important to point out that only $\begin{pmatrix} 0 \\ 1 \end{pmatrix}\begin{pmatrix} 1 \\ 0 \end{pmatrix}$ and $\begin{pmatrix} 0 \\ 1 \end{pmatrix}\begin{pmatrix} 0 \\ 1 \end{pmatrix}$ could result in the imposition of **P (**$-\frac{\pi}{2}$). The first state is the control and the second is the one where the phase-gate is imposed. In this way, the only way to determine whether or not a phase-gate had been imposed, after the target state was processed further, was to measure the control state; in this case, assistant qubits were the control, the others were the target. More explicitly, when q[0] and q[4] were $|1>_{q[0]}$ , $|0>_{q[4]}$ , only one phase-gate was applied, but when they were $|1>_{q[0]}, |1>_{q[4]}$, all three phase-gates were imposed, and the only way to determine the type of SCs that had been imposed was by the measurements of those assistants. Notice in fig. 4 that q[1], q[2], or q[3] had the opportunity to have a constant phase value in each experiment, given that q[0] remained always in the state |1>, allowing control on the phase-gate to always be applied, but the other two qubits, not under the control of q[0], had two possible phase values: **P** $(-\frac{\pi}{2})$ gates were set on both qubits, when the outcome of q[4] was in state |1>, but no **P** $(-\frac{\pi}{2})$ was applied when the outcome of q[4] was |0>. Now, by implementing those phase-gates, with the hadamard gates, the GHZ-like state, in the x-basis

$$|+>=\frac{1}{\sqrt{2}}(|0>+|1>),$$

$$|->=\frac{1}{\sqrt{2}}(|0>-|1>),$$

became

$$|GHZ>'=\frac{1}{\sqrt{2}}(|+++>+e^{i\theta}|--->),$$

where the phase value θ was the sum of the phase values for q [1], q[2], and q[3]. Specifically, when the phase-gates were applied only on one of the qubits, $|GHZ>'$ became

$$|GHZ>^{\mathrm{E}}=\frac{1}{\sqrt{2}}(|+++>+|--->);$$

this is also equivalent to

$$|GHZ>^{\mathrm{E}}=\frac{1}{2}(|000>+|110>+|011>+|101>).$$

Now, when the phase-gates were applied on all of the qubits, the corresponding state was

$$|GHZ>^{\mathrm{O}}=\frac{1}{\sqrt{2}}(|+++>-|--->),$$

which can also be written

$$|GHZ>^{\mathrm{O}}=\frac{1}{2}(|111>+|100>+|010>+|001>).$$

Consequently, when only one phase-gate was applied, measurements on the z-basis gave a non-odd number of |1> states. This is analogous to the GHZ state of three spin-1/2 quanta in the z-basis, where one of them is measured along one orthogonal direction and the other two along another orthogonal direction. In terms of spins, their product would always give a positive number if one can regard the state |0> to be the "up" state (spin $+\frac{h}{2}$) and |1> to be the "down" state (spin $-\frac{h}{2}$). Similarly, the expression for $|GHZ>^{\mathrm{O}}$ implies that when the two phase-gates were applied, together with the other, measurements were an odd number of |1> states, which

is analogous to the case of all measurements of the spins done on the same orthogonal direction where their spin products result in a negative value [25].

Because the outcome of q[4] could be either |0> and |1> , in a way that could not have been known before measurements, the state of q[1],q[2], and q[3] were either $|GHZ>^{E}$ or $|GHZ>^{O}$. The outcome of q[0] was set to be |1>, always imposing a phase-gate on one of the entangled qubits; in contrast, not only the outcomes of q[4] turned all three qubits into state $|GHZ>^{O}$ when the measurements of the H gate on q[4] gave |1>, but also turned them into state $|GHZ>^{E}$ when q[4] was |0>. Notice that q[0] and q[4] were not entangled with q[1], q[2] and q[3], because $|GHZ>^{E}$ and $|GHZ>^{O}$ were not superposed but each one was just associated with a particular outcome of q[4], while q[0] had the same value, making it possible to separate the assistants from the three entangled qubits. In this way, the outcomes from each experiment were two states that could be identified by the measured value of q[0] and q[4]. That is, the states $|GHZ>^{E}$ and $|GHZ>^{O}$ can be identified within the states

$$|E> = ½|00001> + ½|00111> + ½|01011> + ½|01101>,$$

and

$$|O> = ½|11111> + ½|11001> + ½|10101> + ½|10011>;$$

The states $|GHZ>^{E}$ and $|GHZ>^{O}$ are more apparent in states |E> and |O> by rewriting the two expressions using the properties of the tensor product:

$$|E> = |0>(½|000> + ½|011> + ½|101> + ½|110>)|1>$$

$$= |0>|GHZ>^{E}|1>,$$

and

$$|O> = |1>(½|111> + ½|100> + ½|010> + ½|001>)|1>$$

$$= |1>|GHZ>^{O}|1>.$$

The outer qubits were used to identify the type of entanglement obtained. In this way, whenever q[4] was in the state |0>, only the phase-gate **P**$(-\frac{\pi}{2})$ , under the control of q[0], was implemented, making qubits q[1], q[2], q[3] be in state $½|000> + ½|011> + ½|101> + ½|110>$, and in state $½|111> + ½|100> + ½|010> + ½|001>$ if the outcome of q[4] was state |1> which allowed all three qubits to have the same phase-gate **P**$(-\frac{\pi}{2})$. In this way, from the two expressions above, whenever q[0] and q[4] were in states $|1>_{q[0]}$ and $|0>_{q[4]}$, respectively, the expected outcome was $|GHZ>^{E}$; the settings that were generated with those qubits were N-OSC (only one **P**$(-\frac{\pi}{2})$ in one of the three entangled qubits). Likewise, when q[0] and q[1] were in states $|1>_{q[0]}$ and $|1>_{q[4]}$ an OSC was imposed on all the entangled qubits (All with **P**$(-\frac{\pi}{2})$) resulting in $|GHZ>^{O}$. Now, three types of N-OSC were produced by the three experiments shown in fig. 4 simply by changing the location of **P**$(-\frac{\pi}{2})$ under the control of q[0],

i.e., each entangled qubit had the phase-gate one time in all three experiments. Only one type of OSC was produced in each experiment. Consequently, the four types of specific settings in the set

*(8)*

$$\{\boldsymbol{P}_{q[1]}\left(-\frac{\pi}{2}\right)\boldsymbol{I}_{q[2]}\,\boldsymbol{I}_{q[3]}\ ;\ \boldsymbol{I}_{q[1]}\,\boldsymbol{P}_{q[2]}\left(-\frac{\pi}{2}\right)\boldsymbol{I}_{q[3]}\ ;\ \boldsymbol{I}_{q[1]}\,\boldsymbol{I}_{q[2]}\,\boldsymbol{P}_{q[3]}(-\frac{\pi}{2})\ ;\ \boldsymbol{P}_{q[1]}\left(-\frac{\pi}{2}\right)\boldsymbol{P}_{q[2]}\left(-\frac{\pi}{2}\right)\boldsymbol{P}_{q[3]}(-\frac{\pi}{2})\}$$

occurred ($\boldsymbol{P}$ indicates a phase-gate, and $\boldsymbol{I}$ the absence of a phase-gate). Also, notice that an excess of data for $\boldsymbol{P}_{q[1]}\left(-\frac{\pi}{2}\right)\boldsymbol{P}_{q[2]}\left(-\frac{\pi}{2}\right)\boldsymbol{P}_{q[3]}(-\frac{\pi}{2})$ was obtained when compared to each of the other three SCs, given that those occurred only once in each of the three experiment but $\boldsymbol{P}_{q[1]}\left(-\frac{\pi}{2}\right)\boldsymbol{P}_{q[2]}\left(-\frac{\pi}{2}\right)\boldsymbol{P}_{q[3]}(-\frac{\pi}{2})$ occurred in every experiment. However, each experiment produced, $|GHZ>^{\mathrm{E}}$ and $|GHZ>^{\mathrm{O}}$ with the same probability; thus, re-programming the circuit did not "determine" the type of state in each run: the assistant qubits had control of the type of entanglement. In this way, requirements (i) and (ii) (in Sec IB) were satisfied given that |E> and |O> were the result of the random outcomes of q[4] and the constant value of q[0], which necessarily entailed either $|GHZ>^{\mathrm{E}}$ or $|GHZ>^{\mathrm{O}}$; also, requirement (iii) was satisfied given that those states were the result of control phase-gates where the assistant qubits outcomes were the only source of information about the type of setting combinations that corroborated the type of entanglement.

### B. Mermin's inequality for the experiments.

The results from the three experiments were used to calculate the equivalent to the averages of Mermin's polynomial for three quanta, using the expression

*(9)*

$$<\boldsymbol{M_3}>=<\boldsymbol{P}_{q[1]}\left(-\frac{\pi}{2}\right)\boldsymbol{I}_{q[2]}\,\boldsymbol{I}_{q[3]}>\ +\ <\ \boldsymbol{I}_{q[1]}\,\boldsymbol{P}_{q[2]}\left(-\frac{\pi}{2}\right)\boldsymbol{I}_{q[3]}>$$
$$+<\boldsymbol{I}_{q[1]}\,\boldsymbol{I}_{q[2]}\,\boldsymbol{P}_{q[3]}\left(-\frac{\pi}{2}\right)>-<\boldsymbol{P}_{q[1]}\left(-\frac{\pi}{2}\right)\boldsymbol{P}_{q[2]}\left(-\frac{\pi}{2}\right)\boldsymbol{P}_{q[3]}\left(-\frac{\pi}{2}\right)>;$$

the "product" for $|GHZ>^{\mathrm{E}}$ was +1, and —1 for $|GHZ>^{\mathrm{O}}$, if one regards |0> to be +1 and |1> to be —1; under the assumption that those settings were not correlated with the type of HVs in the entanglement, it is possible to conclude that *MI* is also true, from a classical mechanical perspective, for the quantum assisted case without "violation". In other words, given that

$$A1)$$

$$\boldsymbol{F}_{q[\beta]}(n,\mathrm{s}_{q[\beta]})\rightarrow\{+1,-1\}\ such\ that,\mathrm{s}_{\mathrm{q}[\beta]}$$
$$\in\left\{\boldsymbol{P}_{q[\beta]}\left(-\frac{\pi}{2}\right),\boldsymbol{I}_{q[\beta]}\right\},and\ \mathrm{s}_{\mathrm{q}[\beta]}\ \ equals\ either\ \boldsymbol{S}_{q[\beta]}\left(e,t_i^{(\beta)}\right)\ or\ \boldsymbol{S'}_{q[\beta]}\left(a,t_i^{(\beta)}\right),$$

where $\{e\}$ and $\{a)$ are the set of "hidden" variables, in the assistant qubits; $t_i^{(\beta)}$is the time parameter for the $i^{th}$joint measurement ($i^{th}$program "shot"); $\{n\}$ is the set of all "hidden" variables for the entangled qubits,

$A2)$

$\boldsymbol{R}$ is a relation among the "hidden" variables, before the settings were imposed, defined by the expression

$$\{n\}\boldsymbol{R}\{e\} \equiv \{n\} \times \{e\}\ and\ \{n\}\boldsymbol{R}\{a\} \equiv \{n\} \times \{a\},$$

and given that $N_b$ is the number of joint measurements for each experiment b with only one phase-gate on one of the qubits and $N'_b$ correspond to those with a phase-gate on each of the three qubits in the expression

*(10)*

$$< M_3 > = \sum_{b=1}^{3}\sum_{i=1}^{N_b}\frac{1}{N_b}\boldsymbol{F}_{q[b]}\left(n_i, \boldsymbol{S}'_{q[b]}\left(a_i, t_i^{(b)}\right)\right)\cdot\boldsymbol{F}_{q[c]}\left(n_i, \boldsymbol{S}_{q[c]}\left(e_i, t_i^{(c)}\right)\right) \cdot\boldsymbol{F}_{q[d]}\left(n_i, \boldsymbol{S}_{q[d]}\left(e_i, t_i^{(d)}\right)\right) - \frac{1}{3}\sum_{b=1}^{3}\sum_{j=1}^{N'_b}\frac{1}{N'_b}\boldsymbol{F}_{q[b]}\left(n_j, \boldsymbol{S}'_{q[b]}\left(a_j, t_j^{(b)}\right)\right) \cdot\boldsymbol{F}_{q[c]}\left(n_j, \boldsymbol{S}_{q[c]}\left(e_j, t_j^{(c)}\right)\right)\cdot\boldsymbol{F}_{q[d]}\left(n_j, \boldsymbol{S}_{q[d]}\left(e_j, t_j^{(d)}\right)\right)$$

for *b, c, d ∈{1,2,3}, b≠c≠d, c≡b+1 mod 3, d≡b+2 mod 3*, imply that

*(11)*

$$-2\ \leq < \boldsymbol{M}_3 > \leq +2,$$

after all measurements were realized, because there is no correlation between $\{n\}$ and $\{e\}$ (by assumption A2), and thus the setting selection process is independent of the GHZ state; expression (10) is the average of $\boldsymbol{M_3}$ for the three experiments in fig. 4. Also, the same could be obtained by simply writing the average of $\boldsymbol{M_3}$ in terms of the actual SCs, once measurements were performed, and those SCs imposed, treating them as if they have no HVs, and statistically independent from the entanglement, as is the case in the "derivation" of any bell-type of inequality [26]. That is, given four "random hidden" variables $n_h, n_i, n_j\ , n_k\ \in\ \{n\}$ , that occur with the four setting combinations, implies that

*(12)*

$$-2 \le \boldsymbol{F}_1(n_h, \boldsymbol{P}_1(-\frac{\pi}{2})) \cdot \boldsymbol{F}_2(n_h, I_2) \cdot \boldsymbol{F}_3(n_h, \boldsymbol{I}_3) + \boldsymbol{F}_1(n_i, \boldsymbol{I}_1) \cdot \boldsymbol{F}_2(n_i, \boldsymbol{P}_2(-\frac{\pi}{2})) \cdot \boldsymbol{F}_3(n_i, \boldsymbol{I}_3)$$

$$+ \boldsymbol{F}_1(n_j, \boldsymbol{I}_1) \cdot \boldsymbol{F}_2(n_j, \boldsymbol{I}_2) \cdot \boldsymbol{F}_3\left(n_j, \boldsymbol{P}_3\left(-\frac{\pi}{2}\right)\right) - \boldsymbol{F}_1(n_k, \boldsymbol{P}_1(-\frac{\pi}{2})) \cdot \boldsymbol{F}_2(n_k, \boldsymbol{P}_2(-\frac{\pi}{2})) \cdot \boldsymbol{F}_3(n_k, \boldsymbol{P}_3(-\frac{\pi}{2})) \le +2;$$

consequently, for N total “shots” where each SC occurs the same number of times,

*(13)*

$$-2 \le \sum_{h=1}^{\frac{1}{4}N} \left(\frac{1}{N/4}\right) \cdot \boldsymbol{F}_1\left(n_h, \boldsymbol{P}_1\left(-\frac{\pi}{2}\right)\right) \cdot \boldsymbol{F}_2(n_h, \boldsymbol{I}_2) \cdot \boldsymbol{F}_3(n_h, \boldsymbol{I}_3) + \sum_{i=1}^{\frac{1}{4}N} \left(\frac{1}{N/4}\right) \cdot \boldsymbol{F}_1(n_i, \boldsymbol{I}_1) \cdot \boldsymbol{F}_2\left(n_i, \boldsymbol{P}_2\left(-\frac{\pi}{2}\right)\right) \cdot \boldsymbol{F}_3(n_i, \boldsymbol{I}_3) + \sum_{j=1}^{\frac{1}{4}N} \left(\frac{1}{N/4}\right) \cdot \boldsymbol{F}_1(n_j, \boldsymbol{I}_1) \cdot \boldsymbol{F}_2(n_j, \boldsymbol{I}_2) \cdot \boldsymbol{F}_3\left(n_j, \boldsymbol{P}_3\left(-\frac{\pi}{2}\right)\right) - \sum_{k=1}^{\frac{1}{4}N} \left(\frac{1}{N/4}\right) \cdot \boldsymbol{F}_1\left(n_k, \boldsymbol{P}_1\left(-\frac{\pi}{2}\right)\right) \cdot \boldsymbol{F}_2\left(n_k, \boldsymbol{P}_2\left(-\frac{\pi}{2}\right)\right) \cdot \boldsymbol{F}_3\left(n_k, \boldsymbol{P}_3\left(-\frac{\pi}{2}\right)\right) \le +2,$$

which also entails expression *(11)*. Note that each of the terms in the inequality are the average value for a particular setting combination. It is assumed that each average is the same even if different number of “shots” are performed when $N$ is large enough. Thus, the expression for $< \boldsymbol{M_3} >$ may also be written

*(14)*

$$< \boldsymbol{M_3} > = < \boldsymbol{F}_1\left(n_h, \boldsymbol{P}_1\left(-\frac{\pi}{2}\right)\right) \cdot \boldsymbol{F}_2(n_h, \boldsymbol{I}_2) \cdot \boldsymbol{F}_3(n_h, \boldsymbol{I}_3) > + < \boldsymbol{F}_1(n_i, \boldsymbol{I}_1) \cdot \boldsymbol{F}_2\left(n_i, \boldsymbol{P}_2\left(-\frac{\pi}{2}\right)\right) \cdot \boldsymbol{F}_3(n_i, \boldsymbol{I}_3) > + < \boldsymbol{F}_1(n_j, \boldsymbol{I}_1) \cdot \boldsymbol{F}_2(n_j, \boldsymbol{I}_2) \cdot \boldsymbol{F}_3\left(n_j, \boldsymbol{P}_3\left(-\frac{\pi}{2}\right)\right) > - < \boldsymbol{F}_1\left(n_k, \boldsymbol{P}_1\left(-\frac{\pi}{2}\right)\right) \cdot \boldsymbol{F}_2\left(n_k, \boldsymbol{P}_2\left(-\frac{\pi}{2}\right)\right) \cdot \boldsymbol{F}_3\left(n_k, \boldsymbol{P}_3\left(-\frac{\pi}{2}\right)\right) >,$$

avoiding the reference to the number of “shots” when the number for each SC is also large enough. In this way, expression (14) is equivalent to expression (10); in the latter, the number of "shots" was explicitly introduced for each summation, and the average of the three averages for $\boldsymbol{P}_{\boldsymbol{q}[1]}\left(-\frac{\pi}{2}\right)\boldsymbol{P}_{\boldsymbol{q}[2]}\left(-\frac{\pi}{2}\right)\boldsymbol{P}_{\boldsymbol{q}[3]}(-\frac{\pi}{2})$ was introduced given that all three experiments had the same SC as opposed to the others that occurred only once in each experiment.

To calculate the average of $\boldsymbol{M_3}$, the frequency for each computational basis element obtained, after 8000 "shots" for each experiment, was categorized by the type of SCs: when the outer qubits corresponded to a N-OSC ( i.e., $|1>_{q[0]}$ and $|0>_{q[4]}$) and q[0], q[1], q[2] resulted in a N-OE, each count was assigned a number "+1"; however, a number "—1" was assigned to each count with N-OSC when the output corresponded to an OE; the average for each N-OSC in each experiment was calculated with the formula

$$< setting\ combination > = \frac{(N_{non\ odd}\ +\ N_{odd}\ )}{(\ |N_{nonodd}|\ +\ |N_{odd}|)}, \tag{15}$$

where $N_{odd}$ and $N_{non\ odd}$ are the counts for OEs and N-OEs, respectively, obtained by the method described above. Now, when the outer qubits implied an OSC (, i.e., $|1>_{q[0]}$ and $|1>_{q[4]}$), each count was considered to be "—1" if q[0], q[1], and q[2] were indeed an OEs; however, each occurrence of N-OE, with OSCs, were counted as "+1". Because all three experiments had the same OSC, the average was also calculated for each experiment with equation (15), but then the total average of all three average products was calculated. In other words,

$$< total\ OSC > = \frac{1}{3}(< OSC >_1 + < OSC >_2 + < OSC >_3), \tag{16}$$

To calculate MI, the average product for each of the three N-OSC were added together and combined with the average of the three average products of the OSCs, as indicated in expression (10):

$$< \boldsymbol{M_3} > = < N\ OSC >_1 + < N\ OSC >_2 + < N\ OSC >_3 - < total\ OSC > \tag{17}$$

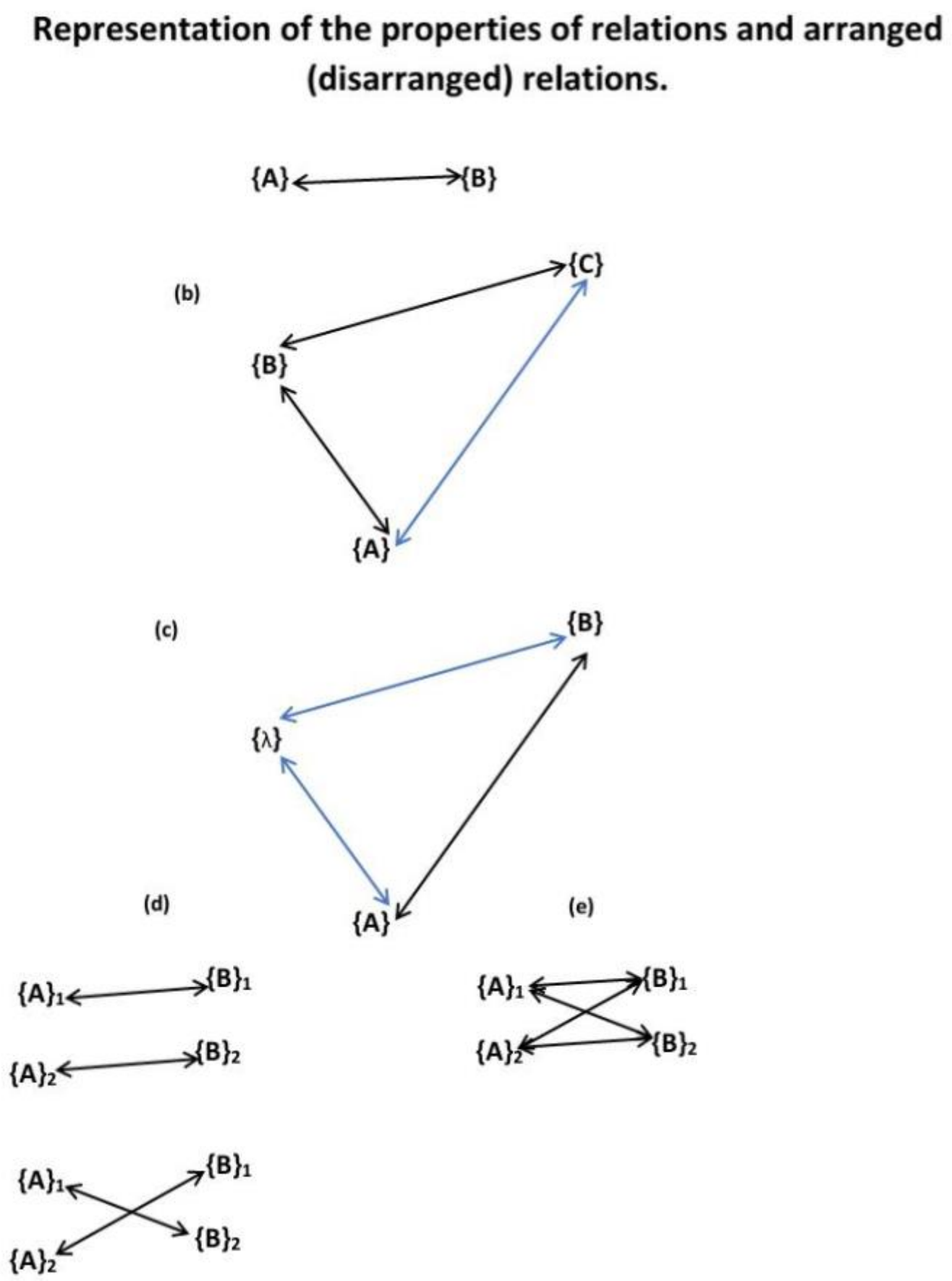

**Figure 8. a.** representation of the symmetric property of **R**. **b.** Representation transitive property of **R,** i.e., the blue edge is the result of the black edges. **c.** Possibly, a relation between two vertices (black edges) can be inferred to be the result of the transitive property between those two vertices and another one (blue edges). **d.** representation of arranged relations (correlations) between two partitions from different sets {A} and {B} **e.** representation of a disarranged relations between two partitions from different sets {A} and {B}.

## C. Replacing the assumptions in Mermin's inequalities.

Assumptions A1 and A2 can be illustrated more clearly with informal relation graphs [27], where the arranged and disarranged relations can be represented (fig. 8d, 8e) as well as the properties of those relations (fig. 8a, 8b). Each vertex of the graph corresponds to an element of a partition (or the partition itself) of a set of HVs; the edges correspond to relations that satisfy,

$$S.P.)\ \{A\}\boldsymbol{R}\{B\} = \{B\}\boldsymbol{R}\{A\}\ and$$

$$T.P.)\ \{A\}\boldsymbol{R}\{B\}\ \&\ \{B\}\boldsymbol{R}\{C\}\ implies\ \{A\}\boldsymbol{R}\{C\},$$

which are the symmetric and transitive property respectively; the properties are represented graphically in fig. 8. It is important to mention that the S.P. and the T.P. are reasonable axioms for the way in which the HVs of a qubit are related to those of another if the HVs are analogous to "reference frame" transformations, i.e., the coordinates in a reference frame A can transform into coordinates in reference frame B and vice versa; also, a coordinate transformation between reference frame A and C can be implied from the transformation that A and C have with a common frame. Also, the same type of graphs can be used to represent the assumptions that are needed to deduce *MI*. Fig. 9 shows the relation graph between the HVs in the assistant qubits and those in the entanglement, where there is a disarranged relation as implied in assumption A2. Notice the impossibility of writing the setting functions in terms of the HVs from the entanglement: two types of settings could be obtained with the same type of entanglement HVs (i.e., there are two paths that lead to the same setting output). Consequently, with the disarranged relations, it is impossible to write the functions of q[0], q[1] and q[2] in terms of the entanglement's HVs in such a way that their output is consistent with the SCs. Thus, the statistical independence of the settings and the HVs of the entanglement is implied by the disarranged relation, which entails further that *MI* should not be "violated".

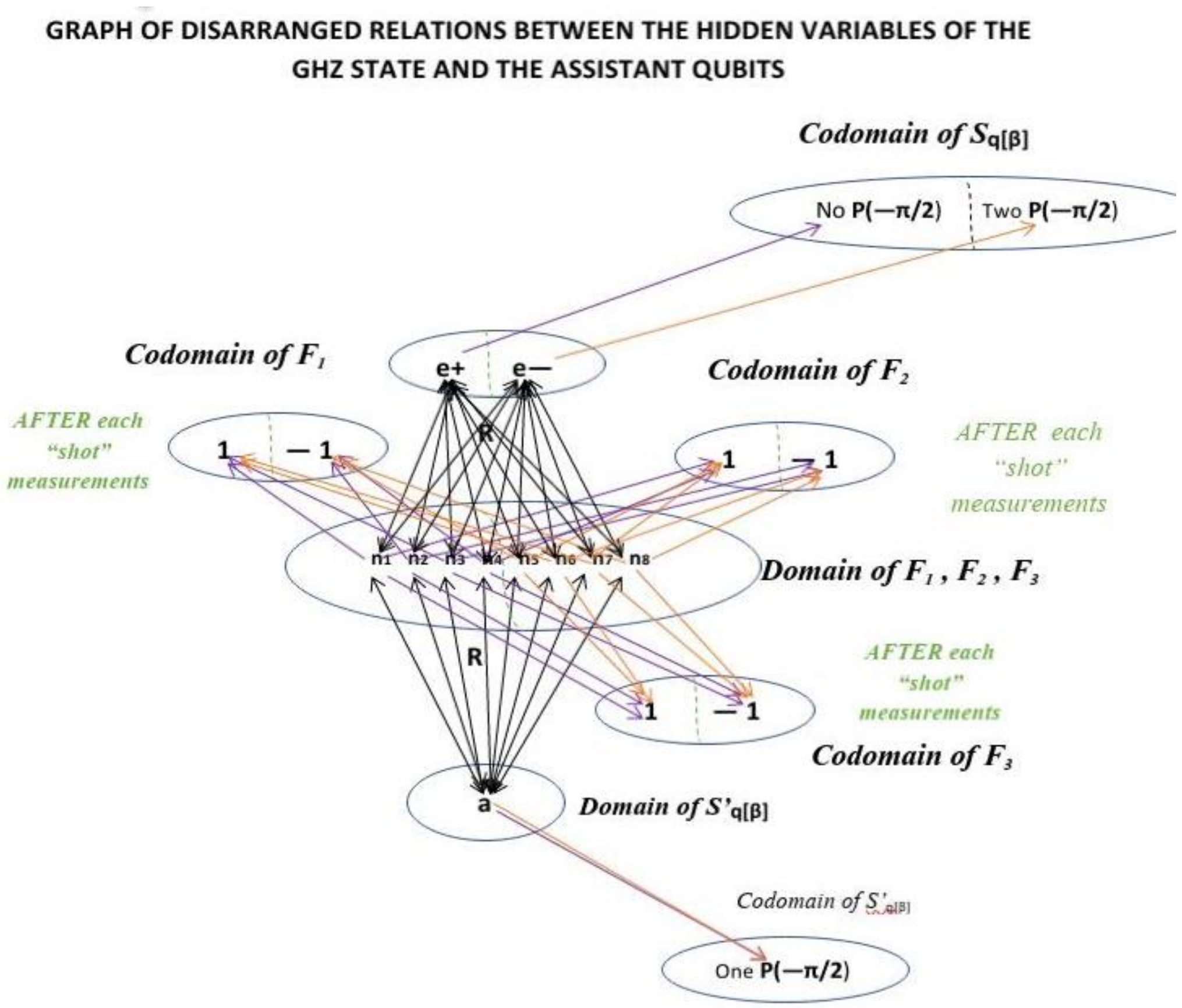

**Figure 9.** The double black arrows represent the relation between set {n} and {e}. The colored arrows indicate the rule that assigns a particular output to every measurement given the "hidden" variable. Notice that there is no arrangement in their "hidden" variables i.e. the same element of {n} has two possible paths that lead to two different setting combinations for the same type of output from the GHZ state.

Assumption *A2* is opened to interpretation in the quantum assisted experiment; it can be replaced by an assumption that regards a statistical interdependence among the HVs before the implementation of the SCs (as implied in the QSD interpretation). Let's call the new assumption *B2*:

$$B2)\ \ Let\ \{n^{(+)}\}\ \boldsymbol{R}\ \{a\},$$

$$\{n^{(-)}\}\ \boldsymbol{R}\ \{a\},$$

$$\{n^{(+)}\}\ \boldsymbol{R}\ \{e^{(+)}\},$$

$$and\ \{n^{(-)}\}\boldsymbol{R}\ \{e^{(-1)}\}$$

be relations that exist before setting on qubits q[1], q[2], and q[3]became defined,

constructed from the set of "hidden" variables $\{n^{(+1)}\} \subset \{n\}, \{n^{(-1)}\} \subset \{n\}, \{e^{(+1)}\} \subset \{e\}, \{e^{(-1)}\} \subset \{e\}$, such that

a) $\{n^{(+)}\}\boldsymbol{R}\ \{a\}, \{n^{(-)}\}\ \boldsymbol{R}\ \{a\}$ are partitions of $\{n\}\ \boldsymbol{R}\ \{a\}$,
b) $\{n^{(+)}\}\boldsymbol{R}\ \{e^{+}\}, \{n^{(-)}\}\boldsymbol{R}\ \{e^{(-)}\}$ are partitions of $\{n\}\boldsymbol{R}\ \{e\}$

$$\Leftrightarrow \begin{Bmatrix} F_{q[b]}\left(n_k^{(+)}, \boldsymbol{S'}_{q[b]}(a_k, t')\right) \cdot\ F_{q[c]}(n_k^{(+)}, \boldsymbol{S}_{q[c]}(e_k^{(+)}, t'))\ \cdot\ F_{q[d]}(n_k^{(+)}, \boldsymbol{S}_{q[d]}(e_k^{(+)}, t')) =\ +1 \\ F_{q[b]}\left(n_k^{(-)}, \boldsymbol{S'}_{q[b]}(a_k, t')\right) \cdot\ F_{q[c]}(n_k^{(-)}, \boldsymbol{S}_{q[c]}(e_k^{(-)}, t'))\ \cdot\ F_{q[d]}(n_k^{(-)}, \boldsymbol{S}_{q[d]}(e_k^{(-)}, t')) =\ -1, \end{Bmatrix}$$

where $b, c, d \in \{1,2,3\}, b \neq c \neq d, c \equiv b + 1\ mod\ 3, d \equiv b + 2\ mod\ 3$, and $t'$ is the time when measurements in all three qubits were done for the $k^{th}$ "shot",i.e., the subscript in each hidden variable indicate a number for each joint measurement. In this way, the domain of $F$ and $S$ are correlated from the start of every "shot" so that their codomains result in two types of patterns. The relation graph for B2 is shown in fig. 10. In contrast to fig. 9, with assumption B2, one path leads from any element of the two partitions of HVs in the GHZ state to a particular setting. In this way, if B2 is true, then

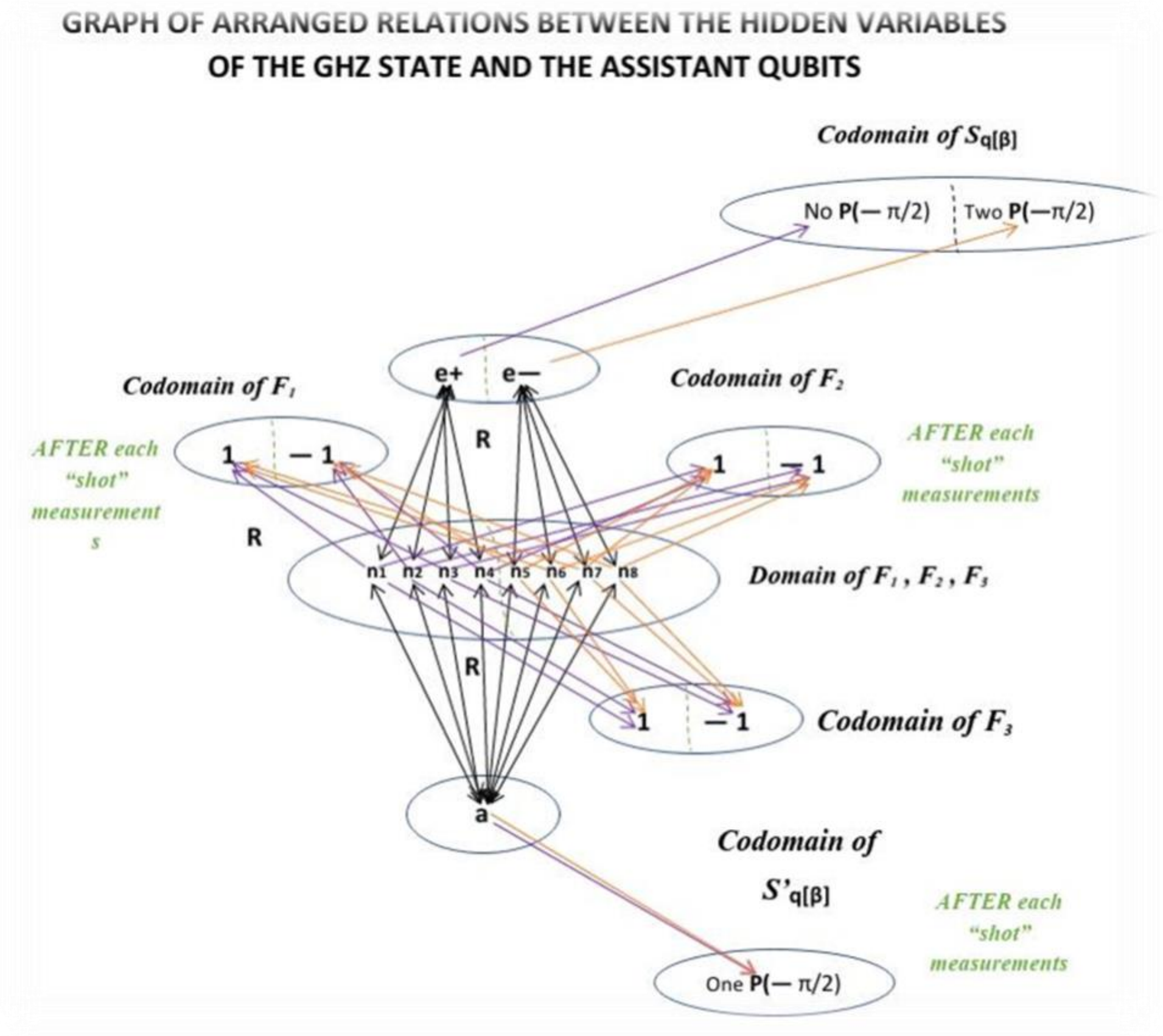

**Figure 10.** The black double arrows show the correlation among the "hidden" variables in the assistant qubits and the GHZ state. Notice that each of the two partitions of their "hidden" variables are connected with their corresponding partition only (indicating an arranged relation). Each path from any of the elements of {n} leads to the correct type of output for both the GHZ state and the assistant qubits.

$$< \boldsymbol{M_3} >_{ideal} = \sum_{b=1}^{3}\sum_{i=1}^{N_b}\frac{1}{N_b}\boldsymbol{F}_{q[b]}\left(n_i^{(+)}, \boldsymbol{S}'_{q[b]}\left(a_i, t_i^{(b)}\right)\right)\cdot\boldsymbol{F}_{q[c]}\left(n_i^{(+)}, \boldsymbol{S}_{q[c]}\left(e_i^{(+)}, t_i^{(c)}\right)\right) \cdot\boldsymbol{F}_{q[d]}\left(n_i^{(+)}, \boldsymbol{S}_{q[d]}\left(e_i^{(+)}, t_i^{(d)}\right)\right) - \frac{1}{3}\sum_{b=1}^{3}\sum_{j=1}^{N'_b}\frac{1}{N'_b}\boldsymbol{F}_{q[b]}\left(n_j^{(-)}, \boldsymbol{S}'_{q[b]}\left(a_j, t_j^{(b)}\right)\right) \cdot\boldsymbol{F}_{q[c]}\left(n_j^{(-)}, \boldsymbol{S}_{q[c]}\left(e_j^{(-)}, t_j^{(c)}\right)\right)\cdot\boldsymbol{F}_{q[d]}\left(n_j^{(-)}, \boldsymbol{S}_{q[d]}\left(e_j^{(-)}, t_j^{(d)}\right)\right) = 4 \tag{18}$$

where $\mathrm{b,c,d} \in \{1,2,3\}, \mathrm{b} \neq \mathrm{c} \neq \mathrm{d}, \mathrm{c} \equiv \mathrm{b}+1 \bmod 3, \mathrm{d} \equiv \mathrm{b}+2 \bmod 3$, for any $\mathrm{t}_{\mathrm{k}}^{(\beta)} = \mathrm{t'}$, such that t' is the time when all measurements were performed for each $k^{th}$ "shot", and β=b,c, or d; the setting functions are the expressions,

(19)

$$\boldsymbol{S'}_{q[\beta]}\left(a_k\,, t_k^{(\beta)}\right) = \boldsymbol{P}_{q[\beta]}\left(-\frac{\pi}{2}\right) for\ \forall t \in (\tau_1, \tau_2)^{a}_{\beta initial} \leq t_k^{(\beta)} \leq \ \forall t \in (\tau_5, \tau_6)^{a}_{\beta measurement},$$

(20)

$$\boldsymbol{S}_{q[\beta]}\left(e_k^{(-)}, t_k^{(\beta)}\right) = \begin{Bmatrix} \boldsymbol{P}_{q[\beta]}\left(-\frac{\pi}{2}\right)\ when\ \ \forall t \in (\tau_3, \tau_4)^{(-)}_{\beta on} \leq t_k^{(\beta)} \leq \ \forall t \in (\tau_5, \tau_6)^{(-)}_{\beta measurement} \\ \boldsymbol{I}_{q[\beta]}\ \ when\ \forall t \in (\tau_1, \tau_2)^{(-)}_{\beta initial} \leq t_k^{(\beta)} \leq \ \forall t \in (\tau_3, \tau_4)^{(-)}_{\beta on}, \end{Bmatrix},$$

(21)

$$\boldsymbol{S_q}_{[\beta]}(e_k^{(+)}, t_k^{(\beta)}) = \boldsymbol{I}_{q[\beta]}\ when\ \forall t \in (\tau_1, \tau_2)^{(+)}_{\beta initial} \leq t_k^{(\beta)} \leq \ \forall t \in (\tau_5, \tau_6)^{(+)}_{\beta measurment},$$

where

$$(\tau_1, \tau_2)^{(+)}_{\beta initial}, (\tau_5, \tau_6)^{(+)}_{\beta measurment}\ , (\tau_3, \tau_4)^{(-)}_{\beta on}, (\tau_1, \tau_2)^{(-)}_{\beta initial},$$

$$(\tau_5, \tau_6)^{(-)}_{\beta measurement}, (\tau_1, \tau_2)^{a}_{\beta initial}, and\ (\tau_5, \tau_6)^{a}_{\beta measurement}$$

are time intervals whose lengths represent time uncertainties during the evolution of the information within the start of each "shot", to the time when measurements where performed. Notice that $< \boldsymbol{M_3} >_{ideal} = 4$ is exactly what is predicted by the quantum theory in a perfect experiment even by the delayed-choice implied in expression (20); however, it is reasonable to expect inefficiencies in the conditions that would make such an ideal number lower due to disarranged relations among the HVs (assumption A2). In other words, assuming that the HVs for a fraction of the total number "shots" were not going to satisfy assumption B2 because of other relations that those HVs might have with the environment (decoherence) [28] in such a way that assumption $A2$ is true for only some of the "shots", the actual expression for the average of $\boldsymbol{M_3}$is given by

*(22)*

$$
\begin{aligned}
< \boldsymbol{M_3} >_{actual} = &\sum_{b=1}^{3}\sum_{k=1}^{N_{b(D)}} \frac{1}{N_{b(D)}} \boldsymbol{F}_{q[b]}\left(n_k^{(v)}, \boldsymbol{S}'_{q[b]}\left(a_k, t_k^{(b)}\right)\right) \cdot \boldsymbol{F}_{q[c]}\left(n_k^{(v)}, \boldsymbol{S}_{q[c]}\left(e_k^{(u)}, t_k^{(c)}\right)\right) \\
&\cdot \boldsymbol{F}_{q[d]}\left(n_k^{(v)}, \boldsymbol{S}_{q[d]}\left(e_k^{(u)}, t_k^{(d)}\right)\right) - \frac{1}{3}\sum_{b=1}^{3}\sum_{l=1}^{N'_{b(D)}} \frac{1}{N'_{b(D)}} \boldsymbol{F}_{q[b]}\left(n_l^{(v)}, \boldsymbol{S}'_{q[b]}\left(a_l, t_l^{(b)}\right)\right) \\
&\cdot \boldsymbol{F}_{q[c]}\left(n_l^{(v)}, \boldsymbol{S}_{q[c]}\left(e_l^{(u)}, t_l^{(c)}\right)\right) \cdot \boldsymbol{F}_{q[d]}\left(n_l^{(v)}, \boldsymbol{S}_{q[d]}\left(e_l^{(u)}, t_l^{(d)}\right)\right) \\
+ &\sum_{b=1}^{3}\sum_{i=1}^{N_{b(A)}} \frac{1}{N_{b(A)}} \boldsymbol{F}_{q[b]}\left(n_i^{(+)}, \boldsymbol{S}'_{q[b]}\left(a_i, t_i^{(b)}\right)\right) \cdot \boldsymbol{F}_{q[c]}\left(n_i^{(+)}, \boldsymbol{S}_{q[c]}\left(e_i^{(+)}, t_i^{(c)}\right)\right) \\
&\cdot \boldsymbol{F}_{q[d]}\left(n_i^{(+)}, \boldsymbol{S}_{q[d]}\left(e_i^{(+)}, t_i^{(d)}\right)\right) - \frac{1}{3}\sum_{b=1}^{3}\sum_{j=1}^{N'_{b(A)}} \frac{1}{N'_{b(A)}} \boldsymbol{F}_{q[b]}\left(n_j^{(-)}, \boldsymbol{S}'_{q[b]}\left(a_j, t_j^{(b)}\right)\right) \\
&\cdot \boldsymbol{F}_{q[c]}\left(n_j^{(-)}, \boldsymbol{S}_{q[c]}\left(e_j^{(-)}, t_j^{(c)}\right)\right) \cdot \boldsymbol{F}_{q[d]}\left(n_j^{(-)}, \boldsymbol{S}_{q[d]}\left(e_j^{(-)}, t_j^{(d)}\right)\right) \leq 4
\end{aligned}
$$

*such that* $v, u \in \{+1, -1\}, v \neq u,$ *and* $N_{b(D)}, N'_{b(D)}$ *are a fraction of the number of "shots", for disarranged relations among the HVs, resulting in non-odd and odd number of "down" states, respectively, for the k and l "shots", in any "imperfect" experiment; a* fraction of $N_{b(A)}, N'_{b(A)}$ *are those for arranged relations, for the i and j "shots";* otherwise, without those disarranged relations, the average for $\boldsymbol{M_3}$ would be exactly equal to four, the expected number by the quantum theory. Also, expression (18) can be written further in terms of just the HVs for the entanglement (without the assistant qubits "frame of reference") given assumption B2; that is,

*(23)*

$$
\begin{aligned}
< \boldsymbol{M_3} >_{ideal} = &\sum_{b=1}^{3}\sum_{i=1}^{N_b} \frac{1}{N_b} \boldsymbol{F}_{q[b]}\left(n_i^{(+)}, s_{q[b]}\right) \cdot \boldsymbol{F}_{q[c]}\left(n_i^{(+)}, s_{q[c]}\right) \cdot \boldsymbol{F}_{q[d]}\left(n_i^{(+)}, s_{q[d]}\right) \\
&- \frac{1}{3}\sum_{b=1}^{3}\sum_{j=1}^{N'_b} \frac{1}{N'_b} F_{q[b]}\left(n_j^{(-)}, s_{q[b]}\right) \cdot F_{q[c]}\left(n_j^{(-)}, s_{q[c]}\right) \cdot F_{q[d]}\left(n_j^{(-)}, s_{q[d]}\right) = 4
\end{aligned}
$$

where *b, c,* and *d* are defined in expression (18), and $s_{q[\beta]}$ $for\ \beta = b, c, or\ d,$ is the actual setting on qubit q[β] as defined in *A1, knowing only that either one is a phase-gate while the others are not, or that all of them are*. It is worth pointing out expression (23) resembles what would be expected in the non-assisted quantum version of the experiment (where a scientist, or a classical device, makes the selection of SCs) except that in this case it is known that the process for setting selection is a function of a quantum superposition, where a measurement implies a particular selection, and *n* is correlated with that process in advanced, but those settings cannot be observed "directly", only implied by the outcome of that superposition. Nevertheless, because *B2* is just an ideal case for all of the data, and some "shots" were going to have disarranged relations (assumption A2), the expected expression is,

*(24)*

$$
\begin{aligned}
< \boldsymbol{M_3} >_{actual} = &\sum_{b=1}^{3}\sum_{k=1}^{N_{b(D)}} \frac{1}{N_{b(D)}} \boldsymbol{F}_{q[b]}\left(n_k^{(v)}, s_{q[b]}\right) \cdot \boldsymbol{F}_{q[c]}\left(n_k^{(v)}, s_{q[c]}\right) \cdot \boldsymbol{F}_{q[d]}\left(n_k^{(v)}, s_{q[d]}\right) \\
&- \frac{1}{3}\sum_{b=1}^{3}\sum_{l=1}^{N'_{b(D)}} \frac{1}{N'_{b(D)}} \boldsymbol{F}_{q[b]}\left(n_l^{(v)}, s_{q[b]}\right) \cdot \boldsymbol{F}_{q[c]}\left(n_l^{(v)}, s_{q[c]}\right) \cdot \boldsymbol{F}_{q[d]}\left(n_l^{(v)}, s_{q[d]}\right) \\
+ \sum_{b=1}^{3}\sum_{i=1}^{N_{b(A)}} \frac{1}{N_{b(A)}} &\boldsymbol{F}_{q[b]}\left(n_i^{(+)}, s_{q[b]}\right) \cdot \boldsymbol{F}_{q[c]}\left(n_i^{(+)}, s_{q[c]}\right) \cdot \boldsymbol{F}_{q[d]}\left(n_i^{(+)}, s_{q[d]}\right) \\
&- \frac{1}{3}\sum_{b=1}^{3}\sum_{j=1}^{N'_{b(A)}} \frac{1}{N'_{b(A)}} \boldsymbol{F}_{q[b]}\left(n_j^{(-)}, s_{q[b]}\right) \cdot \boldsymbol{F}_{q[c]}\left(n_j^{(-)}, s_{q[c]}\right) \cdot \boldsymbol{F}_{q[d]}\left(n_j^{(-)}, s_{q[d]}\right) \\
&\leq 4,
\end{aligned}
$$

such that $b, c, d$ $and$ $N_{b(D)}, N'_{b(D)}, N_{b(A)}, N'_{b(A)}$ are defined in expression (18) and (22); $u$ could be regarded to be "randomly" equal to $+1$ $or$ $-1$. Note that expression (24) is only a function of the HVs in the entanglement and the actual settings (not setting's HVs). Thus, assumption *B2* explains that $<\boldsymbol{M_3}>_{\text{measured}} \geq 2$, as predicted by the quantum theory, but *A1* might still be true due to experimental imperfections. The aim of calculating expression (10) with multiple "shots" is to verify the "violation" of *MI*, which implies that B2 is true for some of the "shots" in expressions (22) or (24), that is, a fraction of the $N_{b(A)}$ and $N'_{b(A)}$ "shots". Thus, MI is not a complete description but it does demarcate quantum from classical information; it is only a sub-case of expression (22) or (24). In this way, one can make the QSD interpretation where not only the HVs of the assistant qubits and the entanglement were correlated BEFORE settings were imposed, but also the disarranged relations were established before processing the "shots", that is, all of the $N_{b(D)}$ and $N'_{b(D)}$ "shots" with a fraction of the $N_{b(A)}$ and $N'_{b(A)}$.

It is worth noting that there are alternates to B2 that have been presented in the literature [29] where a transformation is introduced among the three entangled quanta (which would correspond to qubits in the present experiment). Unlike B2, such a transformation does not regard the HVs to be the result of correlations with the setting selection process, but such a transformation can be applied to ALL experiments that are MGHZM-like, including non-quantum assisted experiments. However, B2 could be generalized to other bell-type of experiments as well when there is quantum assistance, e.g., by establishing **R** among partitions of the HVs of the entanglement and those quantum-assistants, and applying the QSD principles and analysis introduced.

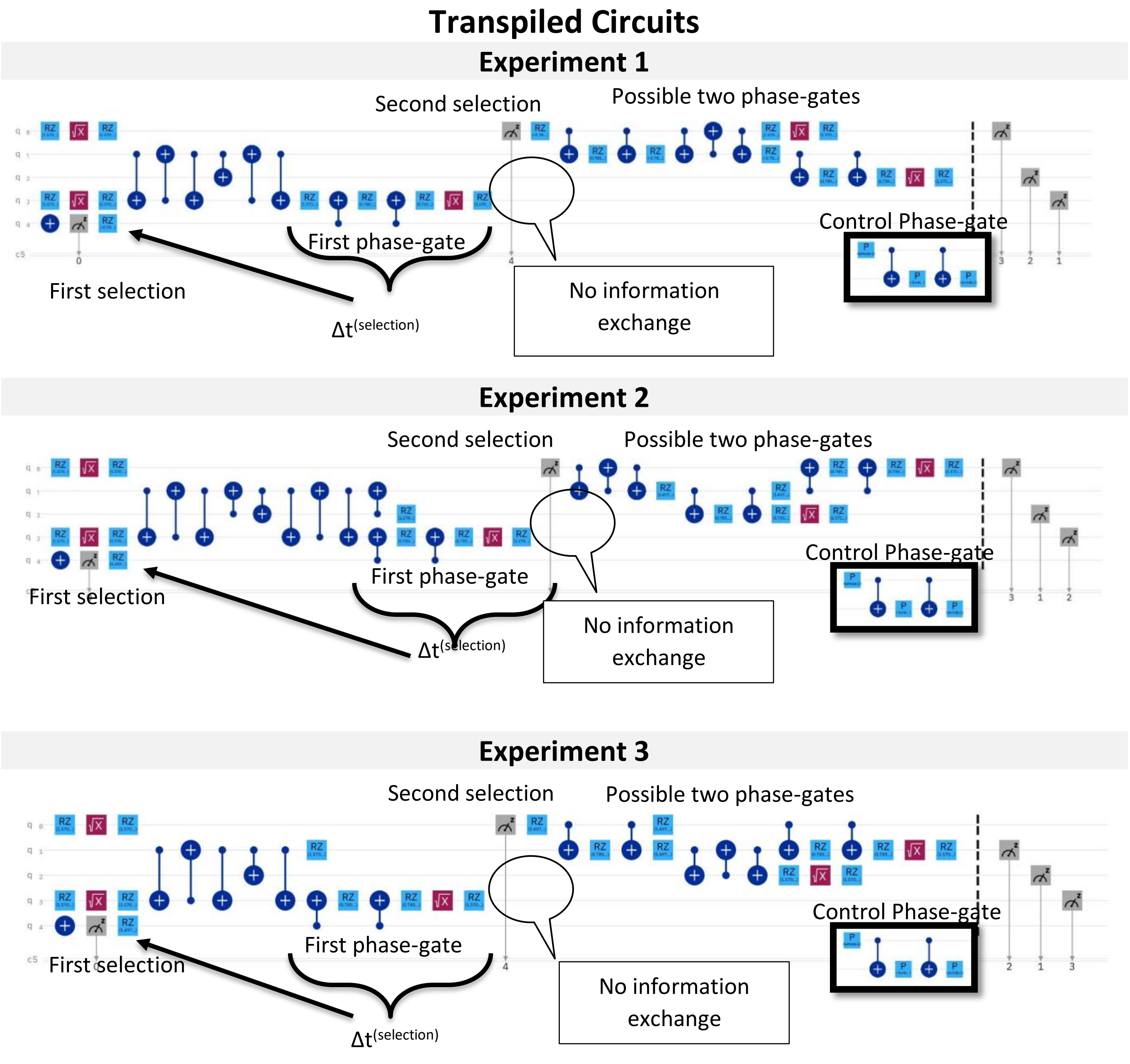

**Figure 11.** The delays between the first and second measurements, for the three experiments, are implied by the number of gates between them. Notice that after the implementation of the first phase-gate on the bottom qubit (corresponding to q[0] in the original circuit) there were no multiple-qubit operations connecting it to the others. Also, the order in which phase-gates were imposed when odd setting combinations occurred is shown; notice the upper qubit (corresponding to q[4] in the original circuit) operated independently of the bottom one, and had control only of the two qubits where the phase-gates were implemented. The side picture helps to identify the control phase-gate in the circuit.

## D. The initial conditions determine the future

One aspect of the experiments depicted in fig. 4, is that the settings for one of the entangled qubits was determined earlier than those for the other two, with unknown time uncertainty, making the SCs "unpredictable" from the start. All the entangled qubits had the opportunity to have their settings being defined first, in each experiment, by one phase-gate. The transpiled circuits for the experiments, shown in fig 11, shows when each assistant qubit was measured relative to the other. Notice that q[0] was measured first (corresponding to q[4] in the transpiled circuit), and a brief gate-time later, q[4] was measured (corresponding to q[0] in the transpiled circuit) before any of the entangled qubits. Their measurements defined the SCs necessarily. The number of gates, in the program process-steps, between the selection of the first phase-gate, and the other two, was defined to be $\Delta t_i^{(selection)}$, for each of the three experiments, by scaling each one to unity (with time uncertainty). Also, the order in which SCs were implemented was the same as the order of selections.

The order in which selection and imposition of settings were done, do not suggest a "classical" interpretation for the arrangement of HVs, DURING each "shot", that can explain what the quantum theory predicts. Notice that q[0], in the transpiled circuit, which corresponds to q[4] in fig. 4, was not functionally related to any of the other qubits, by the two-qubit operations, before it was measured, and this qubit determined whether an ONODS, or not, were going to be obtained.

Furthermore, after the implementation of those two phase-gates, or not, there was no functional relation among the qubits to "communicate" their settings before measurements, as shown in fig. 11 (and fig. 4). Considering qubits to be space-like separated, there is no evidence, from any of the circuits, to indicate that the type of output could be explained through "classical" communication among the qubits during the selection delay or after.

Thus, another possibility is that the HVs can be arranged before each "shot", and the output is determined from the start, as was suggested in the previous sections. Expressions (19), (20), and (21), together with $F_1, F_2, F_3$, imply that there is no need for "retro-causality", as defined in some local models presented in the literature [30], because, in the present case, the future settings depend only on information in the past, assuming that the assistant qubits have an arrangement with the others BEFORE each "shot". Now, there is no evidence, in this case, for a physical system that may have created an arrangement before the start of each "shot", but there is evidence that it is the case whenever there is no interaction among the parts of a correlation in classical systems. Likewise, it is theoretically possible, and reasonable, that those expressions can be time dependent, and that their HVs, BEFORE the start of each "shot", can define what the output was going to be after measurements, that is, assumption B2, as the result of communication paths, BEFORE starting each shot, that may have created the arrangement. In this way, in theory, no matter how much time passes between the first setting imposition and the second, the outcomes correspond to those predicted by the quantum theory and can be interpreted to have been defined from the start by the arrangement among the HVs.

Consequently, although the experiments presented can be considered to be delayed-choice experiments, they do not embrace the same goals of wheeler-type of delayed-choice experiments [31-33], where the future can be interpreted to be able to change the past of a quantum system after measurements, nor there is the intention to interpret the meaning of "erasure" of information with a GHZ-type of entanglement [34]. The present experiments test the ability of the qubits to show correlations, in a sufficiently large number of "shots", even by unexpected variations of the SCs that are under control of two qubits, whose selections are not done necessarily simultaneously, without having evidence for communication paths given the relations among qubits available by the transpiled circuits, that nevertheless can be interpreted with arranged relations before each "shot" like classical systems.

## III. Results

Assuming that the SCs were not correlated with the three entangled qubits' HVs, the results verified "violation" of MI by DCOS; nevertheless, the results can also demonstrate "hidden" variables in all the qubits because the independence between the SCs and the HVs for the entanglement is just another interpretation, and the "violation" of MI could be interpreted to have demonstrated correlation of all the HVs before settings were imposed. The raw data obtained is shown in Histogram 1,2, and 3. By mere inspection of the frequency for each computational basis elements, those corresponding to the terms in |E> and |O> were noticeably more frequent relative to the others (the tallest bars). If the HVs were correlated from the start of the experiment, a fraction of the data in those histograms reflect the frequency for each type of relation; same if there was no correlation. Each term for the expressions of |E> and |O> is indicated underneath each bar (denoting the frequency), with the three middle numbers; the upper qubit corresponds to q[0], in the original circuit in fig. 4, (constant value of "1") and the bottom one corresponds to q[4] (either "1" or "0"). Now, whenever one phase-gate was on any of the three entangled qubits, the upper qubits was "1" and the bottom one was "0"; when all three phase-gates were set, both the upper and lower qubits were "1". The relevant data obtained is shown in table 1. Note that, ideally, the outcomes for q[0] should always be "1"; however, there were some values that returned a "0"; those values were simply discarded from the data; those outcomes did not correspond to any of the SCs necessary to calculate MI with expression (9). In regards to the DCOS, if each gate in the circuit took a unit of time, the transpiled circuits shown in fig. 11 implied a delay between the "selection" of q[0] and q[4]. There were no variations in the delay from each experiment (same gate-time for all three experiments); q[0] (at the bottom of the transpiled circuit, not at the top like in the fig. 4) always was set first in the data presented in tables 1 (which was one reason for discarding the data when q[0] was "0").

## Raw Output

### Experiment 1

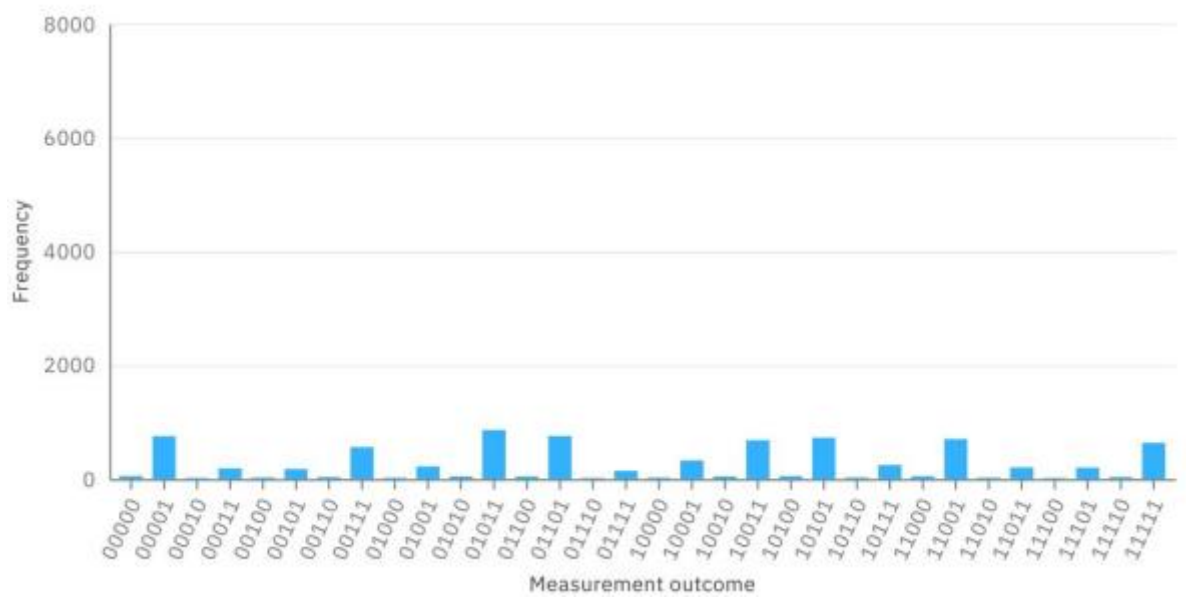

### Experiment 2

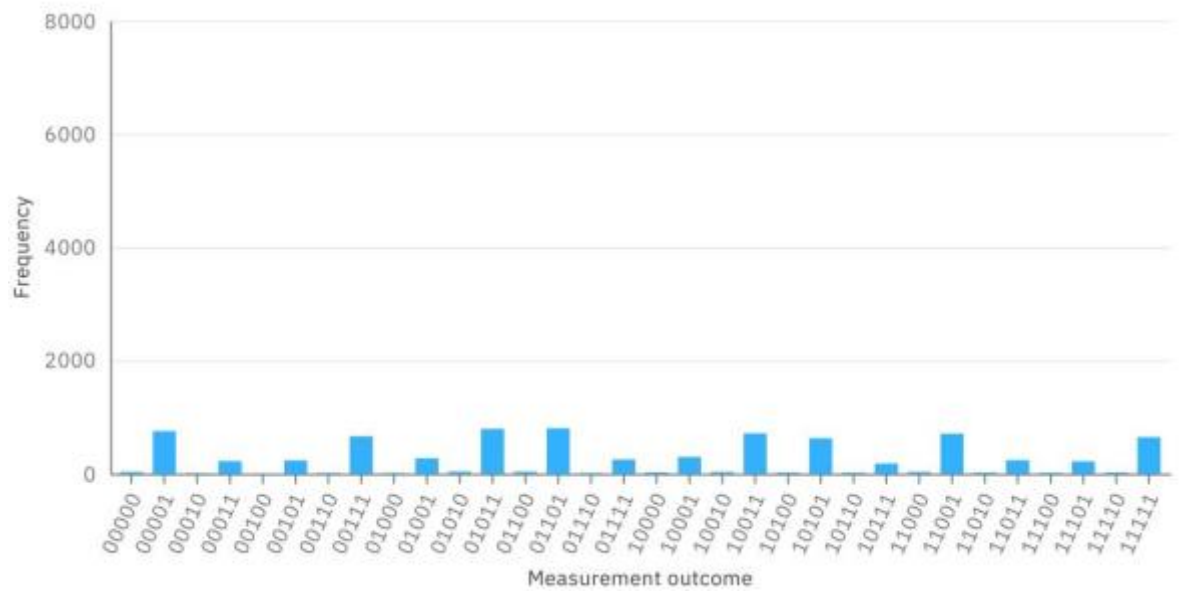

### Experiment 3

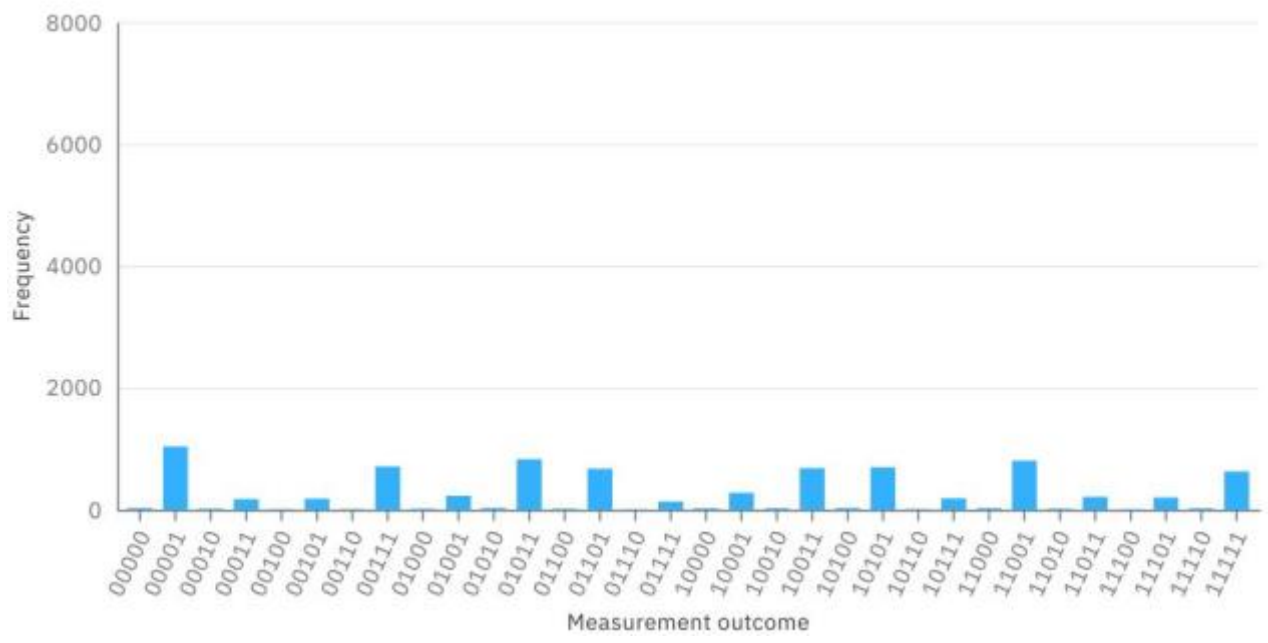

**Histogram 1., 2., 3.** The horizontal axis indicates the computational basis elements. The three middle numbers in each element are the results for the GHZ-like state qubits; outer numbers are the values for the assistants. The vertical axis show the number of joint measurements for each computational basis element. The output was obtained from the IBM quantum computer “Quito”.

From table 1, the calculation for the average of Mermin’s polynomial is straight forward (there is “violation” of Mermin’s inequality).

The histograms 1,2, and 3 can be easily understood by noticing that the left half corresponds to results where the non-odd entanglement can be found, and the right half corresponds to the odd entanglements, where the tall bars are the expected results, and the shorter bars are those that were not the expected given the settings. Now, those other bars that are imperceptible, that appear as spaces between the larger bars, were discarded from the computation of Mermin's inequality; it was only 5.3583333% of the total data and corresponded to different SCs, i.e., two or no phase-gates rather than three or one. Such a data could be attributed to errors in the initialization of q[0]. Now, judging just by the qualities of the larger bars, one can easily see that the frequency of the expected (OSCs matched OEs and N-OSCs matched N-OEs), for both entanglements, was greater than the not expected (OSCs matched N-OEs and N-OSCs matched OEs). Notice also that there is a stronger contrast in the expected/not expected bars on the left side, indicating that the absolute value for each average product of the one-gate-settings was greater than it was for the three-gate-settings, a slight asymmetry that does not follow from the theory, which implied a lower value for the average of Mermin's polynomial.

| Summary of results | | |
|---|---|---|
| EXPERIMENT 1<br>PHASE GATE ON q[1] ONLY<br><br>One P gate<br>Frequencies<br><br>755 -191<br>564 -179<br>863 -224<br>759 -146 TOTAL<br>Sum 2941 -740 3681<br><br>Average Product=<br>0.597935344<br>-------------------------------------------<br><br>THREE PHASE GATES<br><br>Three P gates<br>Frequencies<br><br>-685 328<br>-728 248<br>-704 205<br>-639 202 TOTAL<br>Sum -2756 983 3739<br><br>Average Product=<br>-0.47419096 | EXPERIMENT 2<br>PHASE GATE ON q[2] ONLY<br><br>One P gate<br>Frequencies<br><br>753 -224<br>661 -236<br>792 -273<br>803 -250 TOTAL<br>Sum 3009 -983 3992<br><br>Average Product=<br>0.50751503<br>-------------------------------------------<br><br>THREE PHASE GATES<br><br>Three P gates<br>Frequencies<br><br>-713 296<br>-626 181<br>-708 240<br>-646 219 TOTAL<br>Sum -2693 936 3629<br><br>Average Product=<br>-0.484155415 | EXPERIMENT 3<br>PHASE GATE ON q[3] ONLY<br><br>One P gate<br>Frequencies<br><br>1039 -176<br>710 -182<br>830 -230<br>674 -135 TOTAL<br>Sum 3253 -723 3976<br><br>Average Product=<br>0.636317907<br>-------------------------------------------<br><br>THREE PHASE GATES<br><br>Three P gates<br>Frequencies<br><br>-684 279<br>-696 188<br>-807 212<br>-630 201 TOTAL<br>Sum -2817 880 3697<br><br>Average Product=<br>-0.523938328 |
| $\langle M_3 \rangle$=2.235863182 | | |

**Table 1.** Each of the three columns indicate the three experiments. The top blue-highlighted numbers correspond to the average product for non-odd setting combinations; those bottom blue-highlighted correspond to odd settings. The addition of the three top-highlighted numbers minus the average of those three at the bottom is the value of $\langle \mathbf{M_3} \rangle$.

Now, by looking at the transpiled circuits (Fig. 11), it is possible to assess the relative delay in settings imposed by the outer qubits. Notice that there were seven gates between the measurements of q[0] and q[4] in all three experiment ( which correspond to q[4] and q[0], respectively, in the transpiled circuit).

The quantitative analysis shows a modest violation of Mermin's inequality. The results are summarized in table 1. By adding each of the positive average product totals, for each setting, and subtracting the average for the three average product totals for the negatives, as implied in expression (17), a value of 2.235863182 was obtained. One must keep in mind that these results were determined by the implied settings from the outcomes of the outer qubits but there was no other way to determine the actual settings imposed. However, the slight "violation" implies that expression (22) fits the results, where the disarranged relations correspond to the shorter bars in the histograms ($N_{b(D)}$ $and$ $N'_{b(D)}$ "shots") and the tallest bars are the arranged relations ( $N_{b(A)}$ $and$ $N'_{b(A)}$"shots") combined with a fraction of the disarranged.

Now, the average of the average products of N-OSCs, corresponding to the average of the top row of blue highlighted numbers, in table 1, is greater than the one for OSCs, which is the average of the bottom one. The difference coincides with the fact that the data on the top row were not subject to DCOS like the data on the bottom one, which seems to indicate the deterioration of the results by such a procedure that require control phase-gates (changing settings in "mid-flight") rather than a deviation from the quantum theory.

## IV. DISCUSSION

Although the results exceeded the boundary of MI, the result can be interpreted to affirm that the setting selection process was not independent of the HVs of the entanglement, and thus avoid absolute HVs that exist without reference to the quantum system aiding with the implementation and communication of those settings. Moreover, any joint measurements that did not satisfy the interdependence between the selection process and the entanglement can also be explained by the correlation with other systems in the surroundings; such an explanation would also entail that their arrangement was the result of another with "outside" quantum systems. On the other hand, the question of the locality depends on whether or not both sets of space-like separated HVs can be related to an event they both have in common that can be interpreted to have arranged their HVs in advanced.

**A. Disarranged and arranged relations as a result of the surroundings.**

The results show that MI was not satisfied; thus, a better explanation is a correlation of the HVs before measurements were performed; any discrepancies can also be explained through the same assumption. As shown in expressions (13) and (14), the assumption of the independence of the HVs in the entanglement and the SCs, is used in the "derivation" of MI; because the results did not match the expected, that assumption can be ruled out as the only one. An alternate is expression (22) which correctly predicts the results; thus, a correlation, as stated in *B2*, can be assumed given that they were within the bounds of expression (22). Although results were not exactly equal to four, as it would be the case in a "perfect" experiment, the qubits can also be expected to be related to other systems in the environment in such a way that those relations suppressed correlations in some "shots", assuming that

C1) all HVs exist only in relation to other HVs and

C2) the arranged and disarranged relations among partitions of the HVS from different quantum systems are the result of the S.P. and T.P. of relations that those HVs have with other systems.

Assumption C1 is analogous to the principle of general relativity: just like there are no absolute reference frames to formulate the laws of physics, there are no absolute HVs from which all the results obtained can be formulated. The HVs of the qubits can be expected to have relations with every other quantum system in the surroundings, and those relations also imply the correlations obtained as much as those that were not correlated (as shown in table 1). The way those relations can be deduced from others is described in assumption C2.

There are several ways to obtain disarranged relations among the HVs of the entanglement and the quantum-assistants from the relations those have with the surroundings. For example, given the relations,

$$e1: \ \{\lambda\} \ \mathbf{R} \ \{n\}$$

$$e2: \ \{\lambda\} \ \mathbf{R} \ \{e\}$$

$$e3: \ \{\lambda\} \ \mathbf{R} \ \{a\}$$

such that the left-hand side are the set of HVs for the environment, and assuming that they satisfy the S.P. and the T.P., where each of the sets on the left and the right correspond to HVs from different quanta (qubits), implies that

$$\{n\} \ \mathbf{R} \ \{a\} \ and \ \{n\} \ \mathbf{R} \ \{e\};$$

thus, if one assumes that, in some of the joint measurements, the HVs of both the entanglement and the assistant qubits were related to the same set of HVs from another quantum system, in all possible ways in which two sets can be related, that is, the Cartesian product, then assumption *A1* is implied for those joint measurements. Consequently, those joint measurements related to the HVs of the environment by e1), e2), and e3) would not have "violated" MI if only they had been true for all the joint measurements. Now, here is another example: it is also possible that the HVs from the entanglement and the assistant qubits were uncorrelated with different systems, and also those systems had disarranged relations with each other. In other words, instead of relations e1), e2), and e3), the relations

$$e'1: \; \{\lambda^{(n)}\} \; \mathbf{R} \; \{n\}$$

$$e'2: \; \{\lambda^{(e)}\} \; \mathbf{R} \; \{e\}$$

$$e'3: \; \{\lambda^{(a)}\} \; \mathbf{R} \; \{a\}$$

could be true, such that

$$\{\lambda^{(n)}\}, \{\lambda^{(e)}\}, \{\lambda^{(a)}\}$$

are three sets of HVs for three different quantum systems in the surroundings. If the three sets of HVs are related by

$$e'4: \; \{\lambda^{(n)}\} \; \mathbf{R} \; \{\lambda^{(e)}\},$$

$$e'5: \; \{\lambda^{(n)}\} \; \mathbf{R} \; \{\lambda^{(a)}\},$$

then, assuming T.P. and S.P., follows that

$$\{n\} \; \mathbf{R} \; \{a\} \text{ and } \{n\} \; \mathbf{R} \; \{e\}.$$

On the other hand, the HVs from the entanglement and the assistant qubits can be arranged with HVs from two different systems in the environment that have disarranged relations among themselves; this case also implies that the HVs from the entanglement and the assistant qubits are disarranged in relation to each other; this is illustrated in fig. 12. In other words, given

$$E1^{(+)}: \; \{\lambda^{(+)(n)}\} \; \mathbf{R} \; \{n^{(+)}\},$$

$$E1^{(-)}: \; \{\lambda^{(-)(n)}\} \; \mathbf{R} \; \{n^{(-)}\},$$

$$E2^{(+)}: \; \{\lambda^{(+)(e)}\} \; \mathbf{R} \; \{e^{(+)}\},$$

$$E2^{(-)}: \; \{\lambda^{(-)(e)}\} \; \mathbf{R} \; \{e^{(-)}\},$$

$$E3: \; \{\lambda^{(a)}\} \; \mathbf{R} \; \{a\},$$

*such that* $\{\lambda^{(+)(n)}\}$ *and* $\{\lambda^{(-)(n)}\}$ *are partitions of* $\{\lambda^{(n)}\}$, *and* $\{\lambda^{(+)(e)}\}$ *and* $\{\lambda^{(-)(e)}\}$ *are partitions of* $\{\lambda^{(e)}\}$, one can show that

$$\{n\} \; \mathbf{R} \; \{a\} \text{ and } \{n\} \; \mathbf{R} \; \{e\}$$

If there exist the relations

*RM1:* $\{\lambda^{(n)}\}$ **R** $\{\lambda^{(e)}\}$ *and*

*RM2:* $\{\lambda^{(n)}\}$ **R** $\{\lambda^{(a)}\}$.

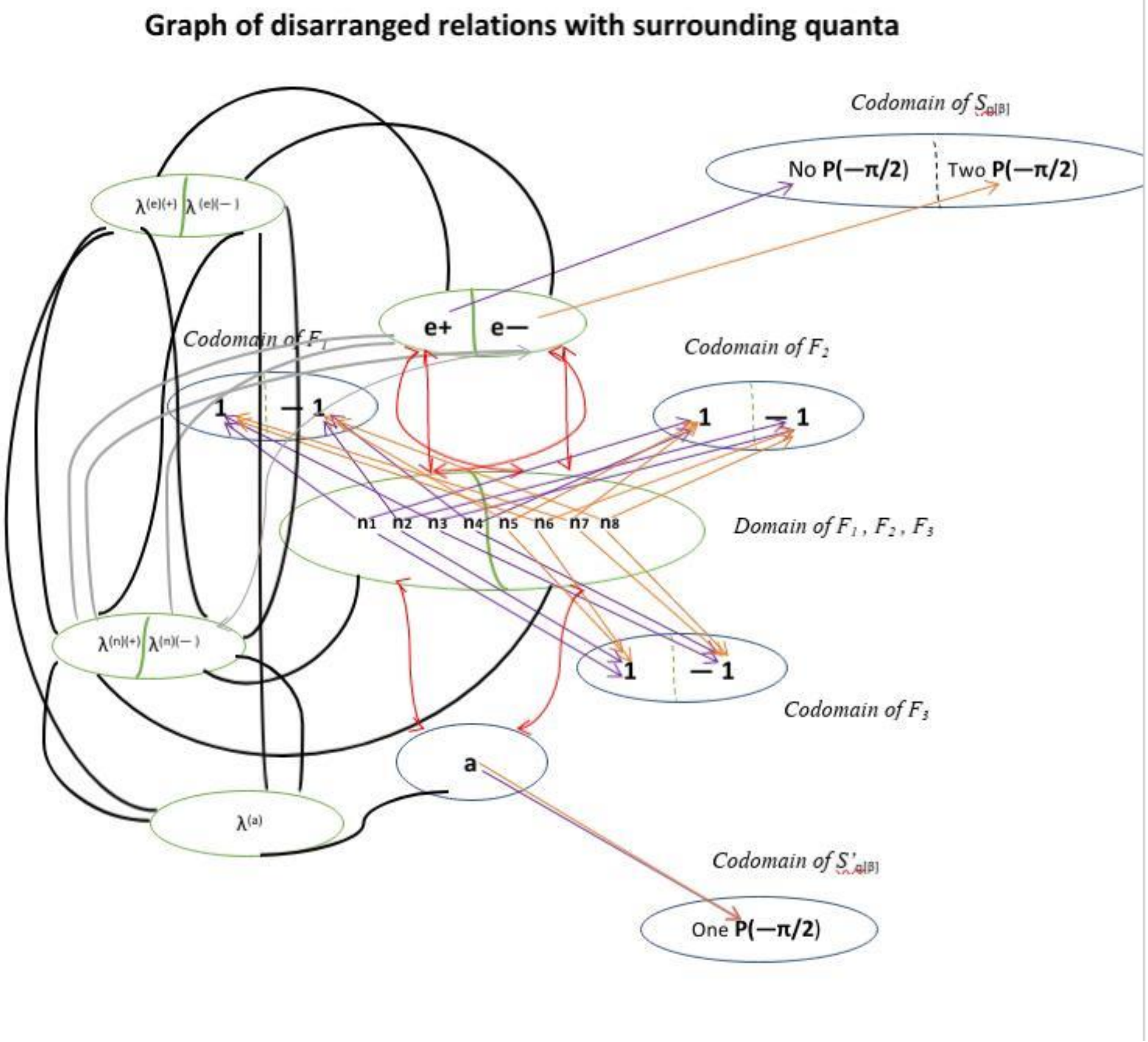

**Figure 12.** Given a disarranged relation between the partitions of $\{\lambda^{(n)}\}$ and $\{\lambda^{(e)}\}$, an arranged relation between {e} and $\{\lambda^{(e)}\}$, an arranged relation between {n} and $\{\lambda^{(n)}\}$, and a relation between {a} and $\{\lambda^{(a)}\}$ (Black edge-bundles), implies a disarranged relation between the partitions for $\{\lambda^{(n)}\}$ and the partitions for {e} (Gray edge-bundles). The latter relations imply further a disarranged relation between {n} and {e}.

The disarranged relations can be "derived" from the graph in fig. 12, applying the T.P. with the graphical method that was shown in fig. 8b., that is, by adding edges to every pair of edges that share only one vertex. In this manner, even disarranged relations can be explained through relations with the environment before measurements. From fig. 12, it is possible to establish an important principle: *the disarranged relations between the partitions of two sets imply also a disarranged one with another set that is related to at least one of the other two if those relations satisfy the symmetric and transitive properties among partitions of HVs for different quanta.*

In this way, the analysis above can also be used to explain correlated relations, and a relationship graph can be used to illustrate it as well; by the same properties of relations, the "violation" of MI can be explained in terms of those that the HVs in the GHZ state and the assistant qubits had with other HVs in the surroundings. For example, let's say that $E1^{(+)}$, $E1^{(-)}$, $E2^{(+)}$, $E2^{(-)}$, E3 are true again, but

$$RM1^{(+)}: \quad \{\lambda^{(+)(n)}\}\ \boldsymbol{R}\ \{\lambda^{(+)(e)}\},$$

$$RM1^{(-)}: \quad \{\lambda^{(-)(n)}\}\ \boldsymbol{R}\ \{e^{(-)(e)}\},$$

$$RM2^{(+)}: \quad \{\lambda^{(+)(e)}\}\ \boldsymbol{R}\ \{\lambda^{(a)}\},$$

$$RM2^{(-)}: \quad \{\lambda^{(-)(e)}\}\ \boldsymbol{R}\ \{\lambda^{(a)}\},$$

are also true instead of RM1 and RM2; it can be shown that the relations between the HVs in the GHZ state and the assistant qubits become

$$\{n\}\ \mathbf{R}\ \{a\},\ \{n^{(+)}\}\ \mathbf{R}\ \{e^{(+)}\},\ and\ \{n^{(-)}\}\ \mathbf{R}\ \{e^{(-)}\},$$

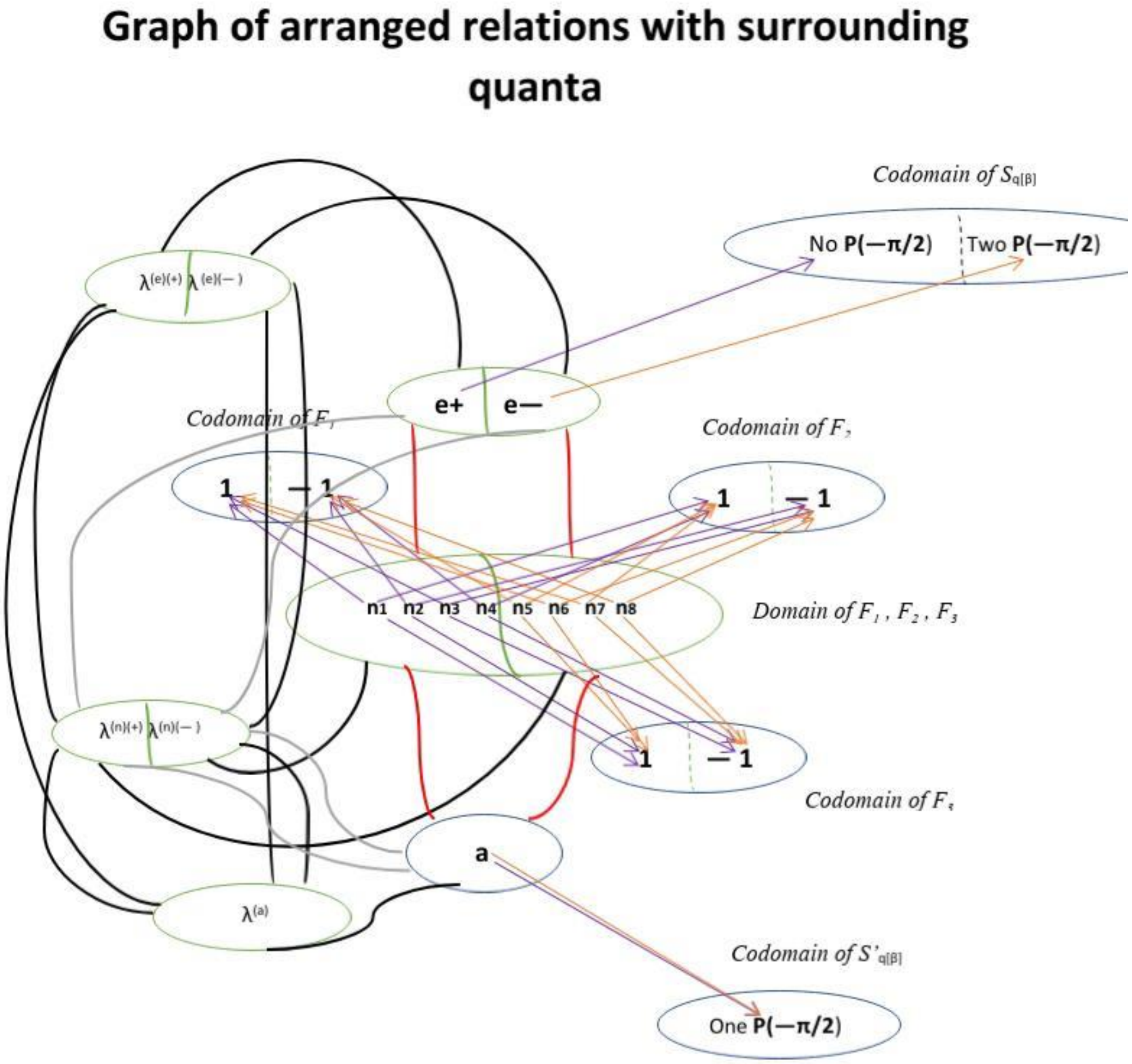

**Figure 13.** Given an arranged relation between the partitions of $\{\lambda^{(n)}\}$ and $\{\lambda^{(e)}\}$, an arranged relation between {e} and $\{\lambda^{(e)}\}$, an arranged relation between {n} and $\{\lambda^{(n)}\}$, and a relation between {a} and $\{\lambda^{(n)}\}$ (Black edge-bundles), implies an arranged relation between the partitions for $\{\lambda^{(n)}\}$ and the partitions for {e} (Gray edge-bundles). The latter relations implies further an arranged relation between {n} and {e} (red edge-bundles)

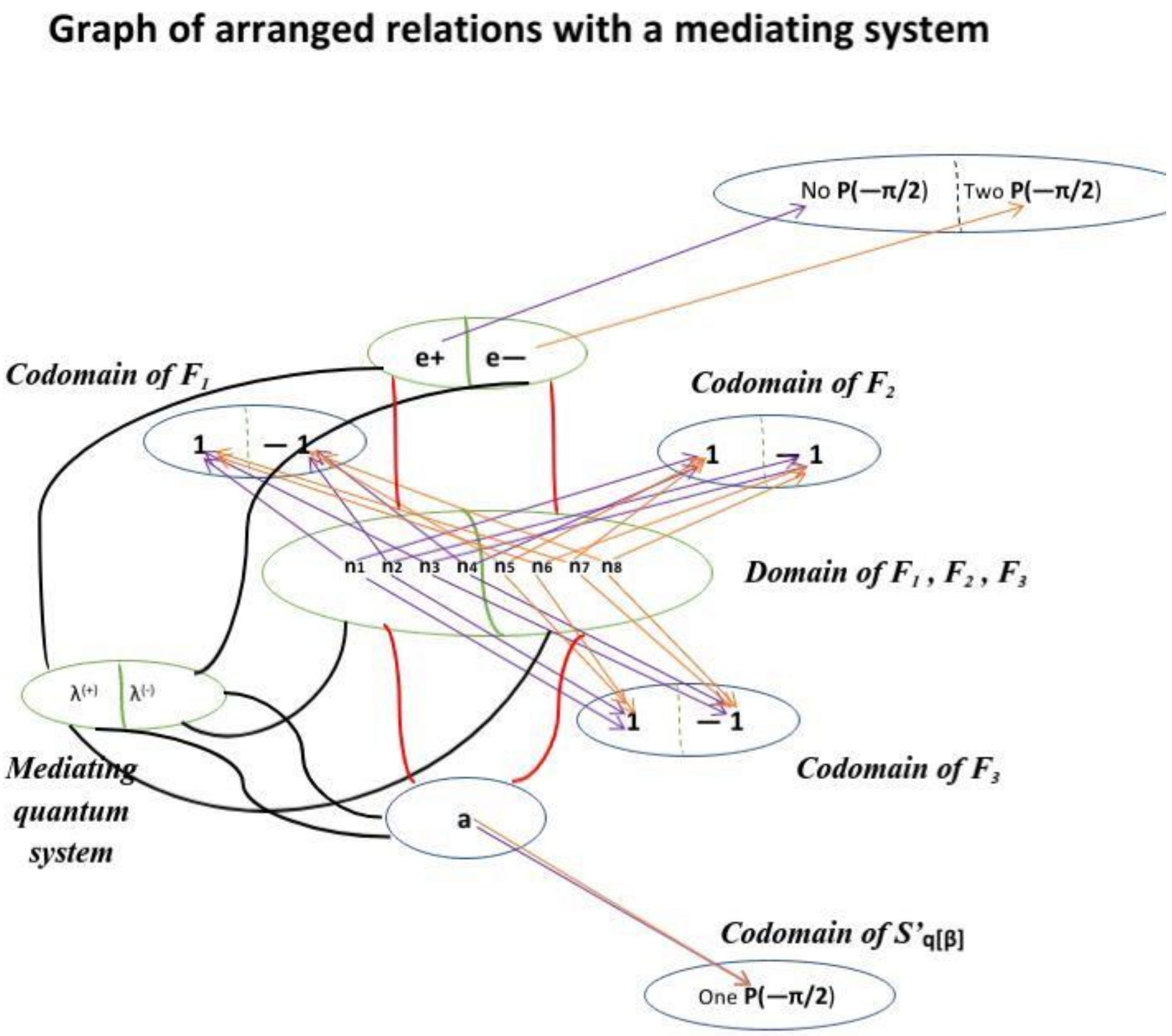

**Figure 14.** The hidden variables of {n} and {e} have arranged relation with {λ} (Black edge-bundles). Because of S.P. and T.P., there is also an arranged relation between {n} and {e}. It is likely that some arranged relations, such as those between {n} and {e}, could be correlated with another set of HVs in common.

by the assumed S.P. and T.P. of relation **R**. To illustrate, fig. 13 now shows the implication of assuming $E1^{(+)}$, $E1^{(-)}$, $E2^{(+)}$, $E2^{(-)}$, E3, $RM1^{(+)}$, $RM1^{(-)}$, $RM2^{(+)}$, $RM2^{(-)}$ with black edge-bundles; in this case, contrasting, the previous example, the relation between the {n} and {e] are arranged (in red) as the result of the replacement of the previous assumptions. Consequently, just like it was expected that some "shots" in the results were not going to satisfy the relations implied in assumption *B2* because of some uncounted relations with surrounding quantum systems, making the average value of $\boldsymbol{M_3}$ be less than four, it is also possible to say that the average of $\boldsymbol{M_3}$ is greater than 2 because of the relations the HVs in all the qubits had with other "unseen" quantum systems before settings were imposed. In this perspective, the quantum mechanical environment defines the relation between {n}, {e}, and {a} (although the latter does not contribute to different types of outputs, but merely affirms that a phase-gate is imposed), as those sets of HVs define the relations between environment HVs, that is, from the red and gray edge-bundles shown in fig. 13, one can also imply the black edge-bundles as shown in fig. 8c. Thus, one may say without formal mathematical rigor, but simple to “see” in the relation graphs by adding edged to every pair sharing just a vertex, to form a closed loop, that

*1) The arranged relation between a pair of quantum systems imply that there is an arranged relation between another pair if the two pairs have an arranged relation (as shown in fig. 13),*

*2) If two quantum systems have an arranged (or disarranged) relation with a third one, those two have an arranged (or disarranged) relation*;

the latter is illustrated in fig. 14 for the case of arranged relations.

## B. Are arranged relations defined by a mediating system before each “shot”?

The principles established above can be extended to the question of whether or not the HVs of the experiments performed are local even though there was “violation” of MI. Notice that the arranged relations, shown in fig. 10, are the best explanation; thus, by adding a mediator to those correlations, one could interpret them to be the origin for the assumed interdependence, BEFORE every joint measurement. Such a mediator could be interpreted to be the cause of the correlations if it is a necessary consequence of the way a superconducting quantum computer initializes the qubits coherently [35-36], in a “simple fiducial state” [37].

In this way, the experiments presented do not confirm that the correlations are local, but it does not rule it out, even though there was “violation’ of MI, if the assumption is that the correlations were established before measurements; further experiments are required to determine whether or not there was a mediating system, BEFORE each joint measurement, that could be interpreted to have provided the arrangements. The principle is the following: *the quantum system in the surroundings that originate the correlation is the mediator that makes the HVs of the correlation local* [30]. The principle is illustrated in fig. 13 and fig. 14; in those relation graphs, one can say that the HVs of the entanglement and the assistant qubits are local; if those surrounding quantum systems could be identified empirically then one could say that those HVs, in fact, are. In this perspective, it is possible that the setting selection process, and the outcomes from the entanglement, can be interpreted to have “emerged” from a common local source [40]. Such a principle is no different from saying, in the QAMGHZM example, that if some device “programmed” the HVs by some local arrangement procedure that relates only specific partitions of the possible HVs with those from the other, all of them did not coordinated themselves as if they are communicating at-a-distance. Also, the principle is implied by the analogy between teleportation of quantum information and the one-time pad cypher [38-39], which is a way of explaining quantum information transfers locally, that is, the information can be conceived to have been encoded "secretly" in both parties, through a common key, and comparison of their information discloses the evidence for the arrangement. On the other hand, if the HVs had just "appeared" in their respective qubits, as if a "quantum potential" [41] defined the relations between them, then the possibility of local HVs can be ruled out; the HVs don't have to be local. Nevertheless, when there is a correlation in classical mechanical systems, where there does not appear to be causal connection between two parts, the cause is looked in other parts in the surroundings. In this way, in the experiments presented, it is possible that a mediator could be identified, that may be interpreted to have had an arranged relations with the assistants as well as the entanglement as if they were classical systems. Because there does not appear to be a clear mechanism during the experiment, it is possible that such a mediator can be identified BEFORE each repetition, given that assumption *B2*. Any quanta that may have had a sufficiently functional relation with the qubits that ultimately initiated the state $|00000>$, can be interpreted to have provided the coordination among the "rules". A toy-circuit that includes an explicit mediating system is shown in fig. 15, corresponding to the circuit version of fig 3. Notice that the circuit, if performed on the quantum computer, could be interpreted in terms of local "hidden" variables only, that is, as if q[0] has control of the HVs for the GHZ state (q[2], q[3], q[4]) and the assistant qubits (q[1] and q[5]), and is able to "program" the proper HVs, on each "shot", so that the "correct" output occurred after measurements. The "choices" of settings that would result from such a toy-model would be local by the same loophole that has been identified in other delayed-choice experiments where a quantum is the ultimate source of information transfers before any selections [42]. The experiments presented do not take advantage of the loophole. What is important about the proposed assumption B2, in the context of the experiments realized in this paper, is that the question of whether or not the implied HVs are local is falsifiable: if there is no local, common quantum system, related to both sets of HVs before each repetition of the experiment, there are no local hidden variables. However, the "violation" of MI does not disprove the existence of local HVs in the quantum assisted version presented.

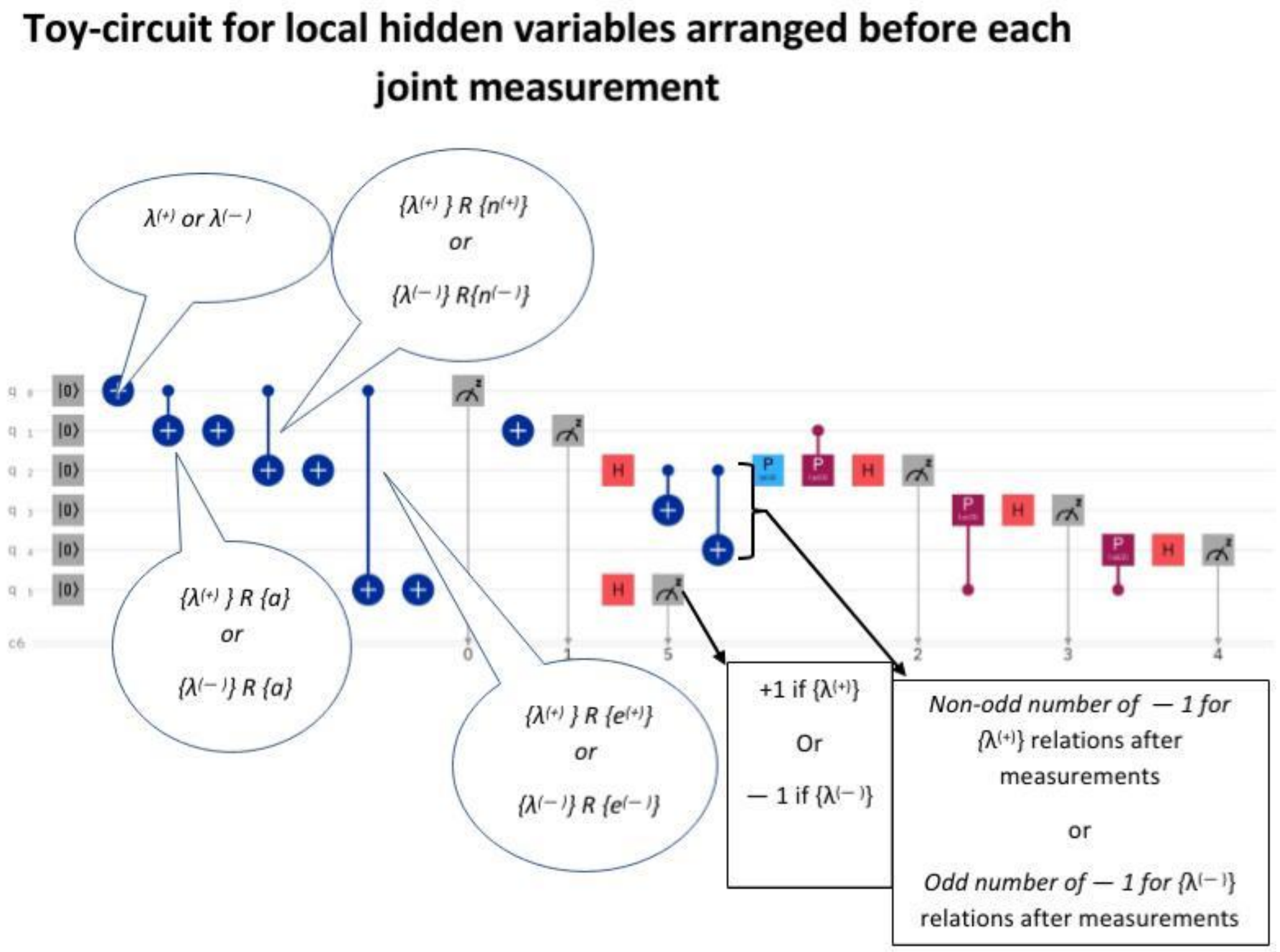

**Figure 15.** q[0] defines the HVs for q[1], q[2] and q[5] with the control gates, which turn all of them into |1> states but then become "flipped" back, ultimately leaving q[1], q[2], q[3],q[4], and q[5] in state the state |0>, before q[1] and q[5] had revealed their "choice" of settings. Because of the relation q[0] had with q[1] and q[5], it is possible to interpret their outcomes as having been "rigged" with those of q[2] which ultimately "rigged" the outcomes with q[3] and q[4]. In this way, the output is determined from the start, by a local distribution of hidden information, even after the information became space-like separated when measurements were performed.

## V. CONCLUSION

The three quantum circuits, that recreated the thought experiment that Mermin presented in "Quantum Mysteries Revisited", demonstrated a technique for evaluating a superconducting quantum computer's ability to maintain an entanglement with unexpected settings caused by other qubits that are not part of that entanglement. However, in this case, the explanation for the "violations" of Mermin's inequality is a set of arranged relation among the partitions of the "hidden" variables for the quantum-assistants and the three entangled GHZ-state qubits BEFORE each "shot"; any joint measurements that did not satisfy the expected quantum mechanical results can be explained by pre-disarranged relations given that the average for Mermin's polynomial was approximately 2.235863182, which exceeded only the "classical" value of 2, but was much less than 4. Those relations can only be assumed when the setting selection process is a function of another quantum system that also informs the type of settings imposed; those relations can be explained by other relations with the surroundings, before each "shot" started, reasonably assuming that all "hidden" variable relations satisfy transitive and symmetric properties. In this way, the hidden variables of the assistants, and the entanglement, could be interpreted to be local if their "hidden" information had been arranged as a result of a common relation with a mediating quantum system, before each "shot", as shown in fig. 13 and 14; such an explanation is falsifiable and requires further research.

# VI. REFERENCES & ACKNOWLWDGEMENTS

I sincerely want to thank Mateus Araujo for his valuable criticisms of quantum superdeterminism posted at https://mateusaraujo.info/2019/12/17/superdeterminism-is-unscientific/ where he articulated clearly the position expounded in the quotation used that was obtained from Ref. [4] below.